\documentclass{aa}  
\usepackage[dvipsnames]{xcolor}

\usepackage{graphicx}
\usepackage{txfonts}
\usepackage{hyperref}
\usepackage[titletoc]{appendix}
\usepackage{subcaption}
\usepackage{soul}
\usepackage{changepage}
\usepackage[normalem]{ulem}
\usepackage{graphicx}
\usepackage{xcolor}
\usepackage[autostyle]{csquotes}
\usepackage{tikz}
\usetikzlibrary{shapes.geometric, arrows}

\begin{document}

   \title{Protoplanetary disk insights from the\\ first ERIS/vAPP survey at 4 $\mu$m}

 \author{F. Maio\inst{1,2},
          V. Roccatagliata\inst{3,1},
          D. Fedele\inst{1},
          A. Garufi\inst{4,8}, 
          A. Zurlo\inst{5,6},
          C. Lazzoni\inst{7},
          S. Facchini\inst{9},
          R.G. Gratton\inst{7},
          D. Mesa\inst{7},
          C. Toci\inst{1,10},
          S. Antoniucci\inst{12,13},
          S. Desidera\inst{7}.
          L. Pino\inst{1},
          E. Rigliaco\inst{7},
          C. Codella\inst{1},
          L. Podio\inst{1},
          V. D'Orazi\inst{14,7},
          G. Lodato\inst{9},
          F. Pedichini\inst{11,12},
          L. Testi\inst{3,13}
          }
   \authorrunning{F. Maio et al.}
   \titlerunning{Protoplanetary disk insights from ERIS/APP survey}

  \institute{
   INAF-Osservatorio Astrofisico di Arcetri, Largo E. Fermi 5, 50125 Firenze, Italy.
              \email{francesco.maio@inaf.it}
         \and
    Università di Firenze, Dipartimento di Fisica e Astronomia, Via Giovanni Sansone 1, 50019 Sesto Fiorentino FI, Italy.
         \and
    Alma Mater Studiorum, Università di Bologna, Dipartimento di Fisica e Astronomia (DIFA), Via Gobetti 93/2, 40129 Bologna, Italy.
         \and
    INAF - Istituto di Radioastronomia, Via Gobetti 101, I-40129 Bologna, Italy.
         \and
    Instituto de Estudios Astrofísicos, Facultad de Ingeniería y Ciencias, Universidad Diego Portales, Av. Ejército 441, Santiago, Chile.
         \and
    Millennium Nucleus on Young Exoplanets and their Moons (YEMS), Santiago, Chile.
         \and
    INAF-Osservatorio Astronomico di Padova, Vicolo dell'Osservatorio, 5, 35122 Padova, Italy.
         \and
    Max-Planck-Institut für Astronomie, Königstuhl 17, 69117 Heidelberg, Germany.
         \and
    Dipartimento di Fisica, Università degli Studi di Milano, Milano, Italy.
         \and
    European Southern Observatory (ESO), Karl-Schwarzschild-Strasse 2, 85748 Garching bei Munchen, Germany.
         \and
    INAF–Turin Astrophysical Observatory, 10025 Pino Torinese (TO), Italy.
         \and
    INAF-Osservatorio Astronomico di Roma, I-00078 Monte Porzio Catone (RM), Italy.
         \and
    ADONI, INAF ADaptive Optics National laboratory of Italy, Italy.
         \and
    Department of Physics, University of Rome Tor Vergata, via della Ricerca Scientifica 1, 00133 Rome, Italy.
    }

   \date{Received **2024; accepted **}

 
  \abstract
   {}
   {We present high-contrast imaging observations of seven protoplanetary disks at 4 $\mu$m using the Enhanced Resolution Imager and Spectrograph (ERIS) on the Very Large Telescope. This study focuses on detecting scattered light from micron-sized dust particles and assessing the potential of the grating vector Apodizing Phase Plat (gvAPP) coronagraph for disk and planet characterization.}
   {Observations were performed in pupil-stabilized mode with the vAPP coronagraph. Data were reduced using reference differential imaging and angular differential imaging techniques, incorporating principal component analysis for point-source detection. Contrast curves and detection limits were computed for planetary companions and disk features.}
   {The infrared disk signal was resolved in all systems, with first-time 4 $\mu$m detections around AS 209 and Elias 2-24, revealing mostly axisymmetric structures extending up to 60 au. Two gaps were detected in the radial profiles of TW Hya (22 au, 35 au) and AS 209 (50 au, 100 au). For Elias 2-24, scattered light emission matched ALMA observations of inner disk structures, marking their first mid-infrared detection. In the case of HD 100546, the vAPP uncovered flared disk structures and faint spiral arms consistent with previous observations. HD 163296 shows a bright inner dust ring, confirming disk asymmetries and features, but we did not detect any planet candidate within the achieved contrast limits. The disk around PDS 70 exhibits clear features, with faint structures detected within the cavity. The observations achieved contrasts enabling the detection of planets down to 800 K, but no companions were detected, implying either low-mass planets, cooler formation scenarios, or a large dust extinction of $A_V\gtrsim20$ mag.}
   {The vAPP performed robustly for imaging structures in protoplanetary disks at 4 $\mu$m, providing critical insights into disk morphology and constraints on planet formation processes. No planetary-mass companions with temperatures >1000 K are present in our sample.}

   \keywords{protoplanetary disks --
                planet-disk interactions --
                protoplanets --
                stars: AS 209, PDS 70, HD 10546, Elias 2-24, HD 163296, TW Hya, HD 169142 --
                techniques: image processing, ADI, RDI --
                instrument: ERIS/NIX, APP
               }

   \maketitle

\section{Introduction}
Protoplanetary disks (PPDs) are the birthplaces of planets, providing the raw materials and environments where these objects form and evolve. The structure and dynamics of these disks, driven by interactions between gas, dust, and embedded bodies, play a central role in shaping the architecture of planetary systems \citep[e.g.,][]{Safronov1969}.
The formation pathways of planets within these disks can vary significantly, with implications for their initial thermal states \citep{Pollack1996}. Planets formed through the  \enquote{hot-start} scenario, typically associated with gravitational instability or rapid accretion of material, are expected to be bright and hot during their early stages, emitting strongly in the infrared \citep[e.g.,][]{Marley2007,2012ApJ...745..174S}. Conversely, the \enquote{cold-start} scenario, often linked to the core accretion model, is characterized by slower accretion and efficient energy dissipation, resulting in planets that are cooler and fainter, making direct detection particularly challenging. However, more recent models \citep{Mordasini2017} suggest that accretion shocks during planet formation can result in significant energy release, leading to higher luminosities even for \enquote{cold-start} planets.  High-contrast imaging provides key measurements of the luminosity of planets during the early stages of their formation, enabling us to obtain insights into the two formation mechanisms.
At the same time, forming planets actively sculpt their natal disks, creating features such as gaps \citep[e.g.,][]{2015ApJ...808L...3A,citazione46,2018ApJ...869L..41A}, cavities \citep[e.g.,][]{2012ApJ...758L..19H,2017A&A...607A..55V}, and spiral arms \citep[e.g.,][]{2012ApJ...748L..22M,2016Sci...353.1519P} that serve as indirect signatures of their presence. Such structures have been detected at multiple wavelengths, from optical/near-infrared (e.g., with the Hubble Space Telescope \citealt{Ardila2007}, and SPHERE at the Very Large Telescope (VLT), e.g., \citealt{2016A&A...588A...8G}) up to submillimeter and millimeter wavelengths with ALMA, VLA, and ATCA \citep[e.g.,][]{2018ApJ...869L..41A,dischi1,Carrasco2016}.

Understanding the interplay between disks and planets requires high-resolution imaging to uncover both the global morphology and fine structures within the disk. Historically, such images have been obtained using instruments capable of high-contrast imaging, with the angular differential imaging (ADI) technique applied to suppress starlight and enhance the contrast of faint structures or companions \citep{2006ApJ...641..556M}.
However, ADI-based methods inherently introduce self-subtraction artifacts, particularly in regions with extended disk emission, making it challenging to reconstruct the disk's intrinsic structure \citep{Milli2012}. In parallel, polarimetric differential imaging (PDI) \citep[e.g.,][]{Gledhill1991,Gledhill2001,Kuhn2001} in the near-infrared (J, H, K bands) with instruments like the Spectro-Polarimetric High-contrast Exoplanet REsearch (SPHERE) has provided insights into the scattering properties of micrometer-sized grains \citep{Benisty2023}. Although these observations are valuable for tracing the disk's surface layers, they are insensitive to optically thick regions or the emission from embedded planetary companions.

Recent advancements in instrumentation and observational techniques have opened up new windows for studying disks in the mid-infrared. Observations at longer infrared wavelengths (3-5~$\mu$m) are particularly advantageous for several reasons. Longer wavelengths reduce scattering and opacity effects, allowing observations to trace deeper layers into the disk and reveal regions hidden at shorter wavelengths \citep[e.g.,][]{Garufi2017,2020MNRAS.492.3440S}. Finally, the L band is ideal for detecting young, forming planets that are still hot and luminous, offering the potential for us to identify companions embedded within the disk’s densest regions \citep[e.g.,][]{2020MNRAS.492.3440S}.

This study focuses on characterizing disk structures through direct imaging at 4~$\mu$m. At this wavelength, the disk signal is dominated by stellar light scattered by micron-sized dust particles. Observations were conducted using the newly commissioned Enhanced Resolution Imager and Spectrograph 
 \citep[ERIS;][]{2023A&A...674A.207D} at the VLT, equipped with the Apodizing Phase Plate (APP) coronagraph. A high-contrast imaging survey of seven disks was performed to explore the potential of this approach.\\
 This paper is organized as follows. Section 2 describes the observations with VLT/ERIS, including the details of the vAPP coronagraph. Section 3 presents the data reduction and image processing techniques. The main results, including disk structures and point-source candidates, are discussed in Section 4. We compare our findings with previous studies and discuss implications in Section 5. Finally, conclusions are provided in Section 6.

\begin{table*}[h]
\centering
\caption{Summary of the main characteristics of the targets in this survey, including celestial coordinates, distance, L' magnitude, system age, and disk inclination ($i$).}
\begin{tabular}{ccccccccc}
\hline
\textbf{Target} & \textbf{RA [J2000]} & \textbf{DEC [J2000]} & \textbf{L'[mag]} & \textbf{dist. [pc]} & \textbf{ Age [Myr]} & \textbf{$i$ [$^{\circ}$]} & \textbf{Sp. Type} & \textbf{refs}\\
\hline
\hline
HD 100546 & 11:33:25.44 & -70:11:41.24 & 3.95 $\pm$ 0.27 & 108.1 $\pm$ 0.4 & 5.0 $\pm$ 1.3 & 42 & A0V & 7,1,8,9 \\
TW Hya & 11:01:51.90 & -34:42:17.03 & 6.97 $\pm$ 0.10 & 60.1 $\pm$ 0.1 & 6.3 $\pm$ 3.7 & 7 & K6V & 7,1,12 \\
PDS 70 & 14:08:10.15 & -41:23:52.58 & 7.93 $\pm$ 0.03 & 112.4 $\pm$ 0.2 & 8.0 $\pm$ 1.0 & 52 & K7IV & 3,4,5,6 \\
AS 209 & 16:49:15.30 & -14:22:08.64 & 6.22 $\pm$ 0.08 & 121.2 $\pm$ 0.4 & 6.9 $\pm$ 0.4 & 35 & K4V & 7,12 \\
Elias 2-24 & 16:26:24.09 & -24:16:13.30 & 5.87 $\pm$ 0.64 & 139.2 $\pm$ 1.2 & $\sim$ 0.4 & 29 & 	K6 & 10,11 \\
HD 163296 & 17:56:21.29 & -21:57:21.88 & 3.46 $\pm$ 0.44 & 101.0 $\pm$ 0.4 & 7.1 $\pm$ 0.6 & 47 & A3V & 7,12 \\
HD 169142 & 18:24:29.78 & -29:46:49.33 & 6.00 $\pm$ 0.10 & 114.9 $\pm$ 0.4 & 12.3 $\pm$ 6.4 & 13 & F1V & 1,2 \\
\hline
\end{tabular}
\tablebib{1. \cite{citazione13}, 2. \cite{citazione46}, 3. \cite{citazione9}, 4. \cite{citazione26}, 5. \cite{citazione29}, 6. \cite{citazione38}, 7. \cite{citazione2}, 8. \cite{citazione14}, 9. \cite{citazione27}, 10. \cite{dischi1}, 11. \cite{dischi2}, 12. \cite{citazione34}}
\label{tab:combined}
\end{table*}

\section{Observations}

The observations of the whole sample were obtained with VLT/ERIS during the night of May 28, 2023 (summary in Table \ref{tab:datistelle}, Program ID 111.24LT.001). We performed high-contrast imaging observations in pupil stabilized mode using the grating vector Apodizing Phase Plate (gvAPP, referred to as vAPP from now on) coronagraph \citep[e.g.,][]{Snik2012, 2021ApOpt..60D..52D,2023A&A...674A.207D}. We adopted the narrow-band filter Br-$\alpha$-cont ($\lambda_{c}$ = 3.965 $\mu m$ , FWHM = 0.108 $\mu m$). Details and atmospheric conditions of the observations are presented in Table (\ref{tab:datistelle}). 
The weather conditions were rather unstable; toward the end of the night, the cloud cover increased, significantly affecting the observations of HD 169142. The DIMM seeing was slightly above 1\farcs0 during the night, except for a peak around 04:30 UT exceeding 2\farcs0. All data related to the observed targets are reported in Table \ref{tab:combined}.

\subsection{Observations structure}
To properly characterize and subtract the thermal background, we carried out an ABBA nodding sequence and we stored the frames in cube mode (i.e., saving each individual frame).
The observations were obtained in full frame mode for all targets, except for HD 100546, which was observed in the upper subsection of the detector (window 4).
The last target observed during the night was HD 169142, which was observed but could not be fully reduced due to the presence of clouds causing an unstable background over time.\\
The associated dark calibration frames were taken during daytime calibration. The flat-field frames were taken on sky during twilight using the broad-band filter L'.

\begin{figure}[t]
        \centering
        \includegraphics[width=0.45\textwidth]{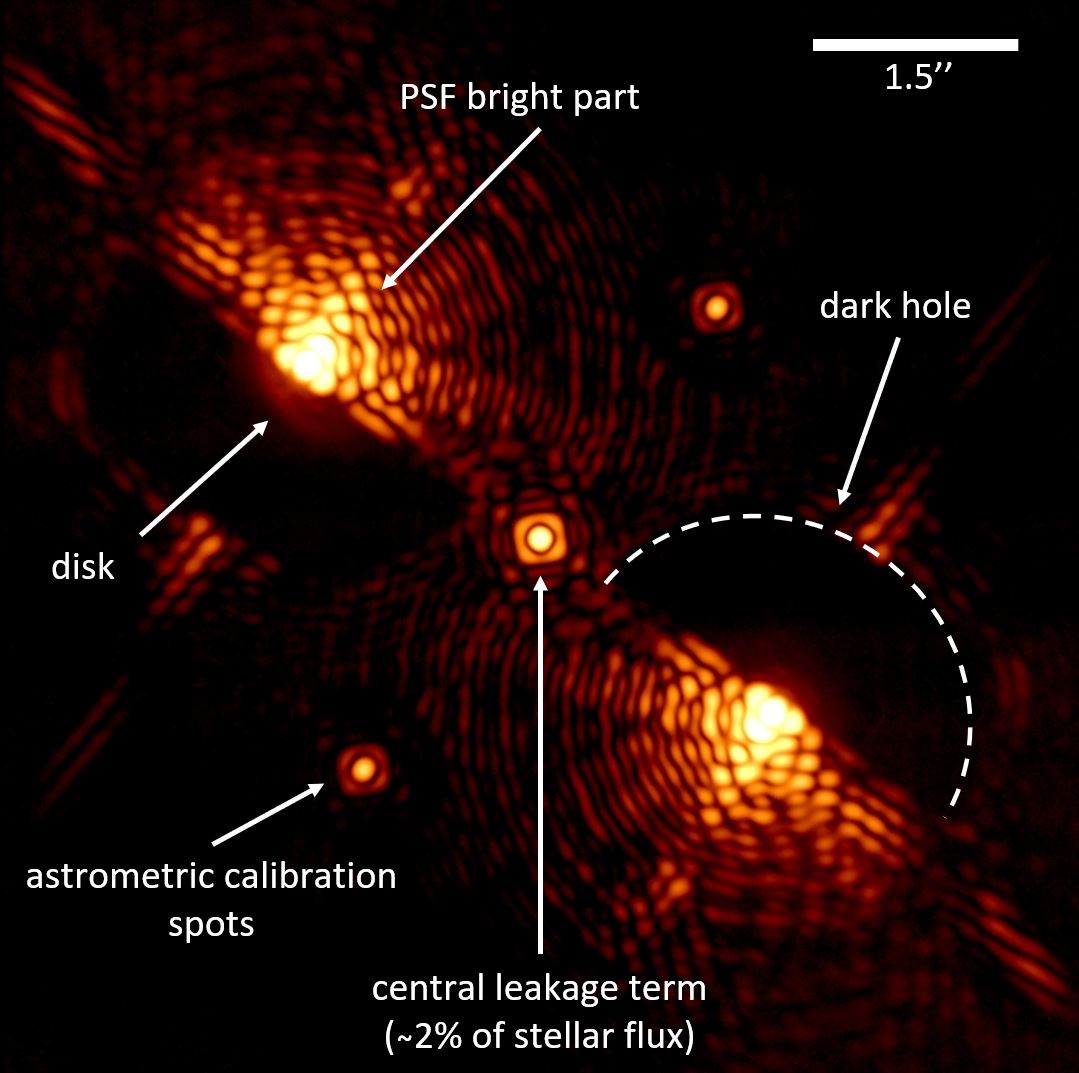}
    \caption{Image of the ERIS vAPP PSF of the star HD 163296. The \enquote{upper PSF} is recognized at the top with its dark hole pointing downward and the bright side at the top. In the dark hole, the presence of the PPD can be observed. The same applies to the \enquote{lower PSF,} but without the disk. In the center, located between the two PSFs, is the \enquote{leakage-term.} The last two points are the \enquote{astrometric calibration spots.}}
   \label{img:APP-PSF}
\end{figure}

\subsection{ERIS Apodizing Phase Plate coronagraph} \label{APP presentazione}
The ERIS APP, located in the pupil plane, modifies the incoming wavefront's complex field by adjusting the phase, resulting in a D-shaped dark region (dark hole). This method provides a higher throughput compared to traditional amplitude apodizers. 
ERIS is equipped with a 180-degree APP, specifically a grating vector APP, which employs a half-wave retarder with an orientation pattern encoded in its fast axis, achieving high-contrast performance across a broad wavelength range. Unlike traditional APPs, which create a dark hole on only one side of the field of view and require two observations with the mask rotated by $180^\circ$ to achieve full coverage, the gvAPP (hereafter vAPP) generates complementary dark holes on both sides of the field in a single observation \citep{Otten2017}. An example of the ERIS image of the star HD 163296 is shown in Figure \ref{img:APP-PSF}.\\

The ERIS APP's point spread function (PSF) is characterized by five main components: 
\begin{itemize}
    \item The upper and lower PSFs, divided into the bright side where the PSF of the star is not suppressed and the dark hole, where the PSF of the star is suppressed. According to the User Manual, these two structures should have about 49\% of the total flux of the star each.
    \item The central leakage-term, with about 2\% of the star's total flux. It has the same shape as the PSF of the telescope. This point marks the true position of the star in the field.
    \item The two astrometric calibration spots were used to improve the astrometric precision of the image.
\end{itemize}

The characterization of the PSF is presented in detail in Appendix \ref{PSFcarat}.

\begin{table*}[ht]
    \centering
    \caption{Summary of the observational characteristics for each target, including observation filter, estimated seeing (from ESO weather report), wind speed, total field rotation, single DIT, total on-target time (ToT), and the quantity of discarded frames due to insufficient instantaneous seeing.}
    \begin{tabular}{cccccccc}
    \hline
    \textbf{Target} & \textbf{Seeing ["]} & \textbf{Wind [m/s]}& \textbf{ $\Delta$PA [$^{\circ}$]} & \textbf{DIT [s]} & \textbf{ToT [min]} & \textbf{Rejected frames} & \textbf{Observation Time (UT)} \\ \hline \hline
    { HD 100546} & { 1} &{7.7}& { 17} & { 3} & { 40} & { 9} & { 22:38 to 23:59} \\
    { TW Hya} & { 1} & {7.8}&{ 26.1} & { 3} & { 56} & { 1} & { 00:00 to 01:20} \\
    { PDS 70} & { 1} & {7.3} & { 38} & { 4} & { 48} & { 0} & { 01:20 to 02:25} \\
    { AS 209} & { 1.3} &{10.1}& { 57.6} & { 3} & { 56} & { 51} & { 03:41 to 05:07} \\
    { Elias 2-24} & { 1.5} &{9.2}& { 3.8} & { 3} & { 28} & { 45} & { 05:30 to 07:08} \\
    { HD 163296} & { 1} &{7.0}& { 1.2} & { 3} & { 56} & { 35} & { 07:09 to 08:38} \\
    { HD 169142} & { 1} &{4.7}& { 1} & { 3} & { 30} & { 8} & { 08:38 to 09:45} \\ \hline
    \end{tabular}
    \label{tab:target_characteristics}
\label{tab:datistelle}
\end{table*}

\section{Data reduction and Image processing}
The data reduction was performed using a custom pipeline, specifically developed for the ERIS/vAPP coronagraphic observations. The main steps of the reduction are: dark subtraction; flat field normalization; bad pixel correction (Sec. \ref{Calibration}); background subtraction (Sec. \ref{Background subtraction}); PSF alignment (Sec. \ref{Alignment}); - frames selection (Sec. \ref{Frames selection}) and PSF subtraction (Sec. \ref{Residual PSF subtraction}). A schematic version of the pipeline can be seen in Appendix \ref{pipelinescheme}.

\subsection{Calibration }\label{Calibration}
\begin{figure*}[!t]
        \centering
        \includegraphics[width=1\textwidth]{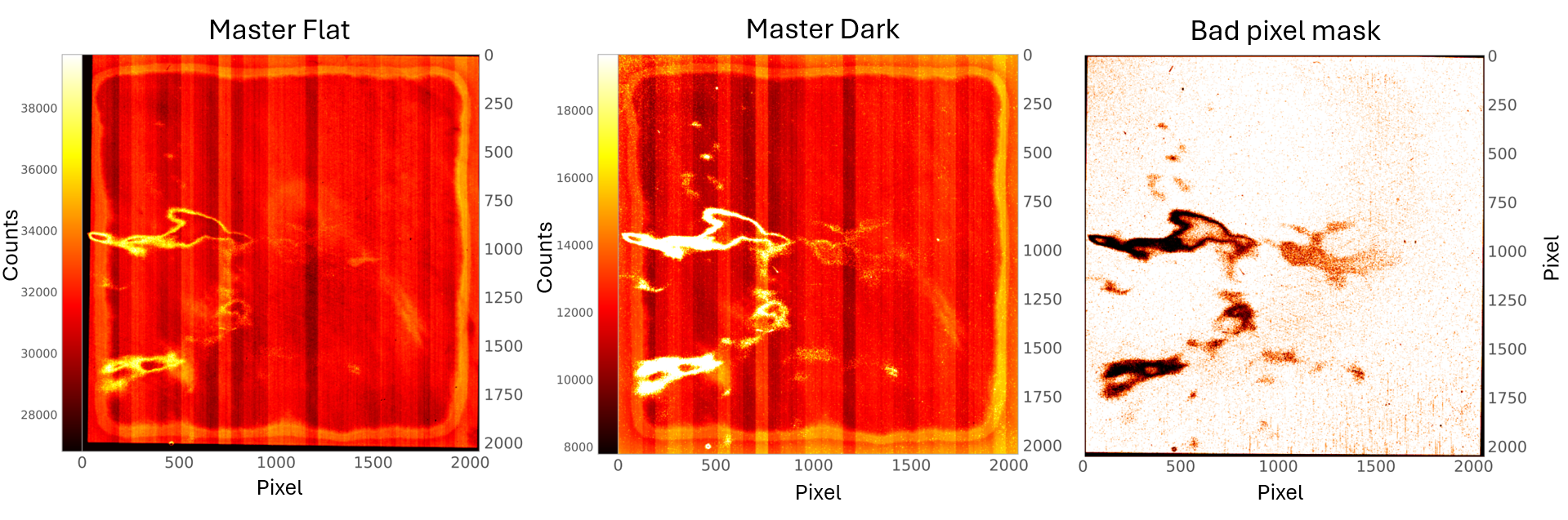}
    \caption{a) Master flat frame. b) Example of a master dark. c) Bad pixel mask where larger concentrations are represented in black.}
    \label{dark-confrnto}
\end{figure*}

The first step was the creation of the master dark frame as the average of all five individual dark frames taken with the same detector integration time (DIT) of the science frames.
This minimizes the statistical effect of thermal noise and averages out temperature variations. An example of the master dark can be seen in Figure \ref{dark-confrnto}.b.

The sky flat-field images were taken in the $L'$ filter and recorded on the 
night after of the observation, during twilight. We used five flat-field images taken at an elevation of 35 degrees, similar to the science frames.
Although differential vignetting and optical aberrations can introduce systematic errors when calibrating observations with flat-fields taken in a different filter, their impact in this case was minimal. The central wavelengths of the $L'$ flat-field filter and the Br-$\alpha$-cont observation filter were very close, meaning that wavelength-dependent optical distortions and throughput variations are negligible.
After the subtraction of the master dark, we created the master flat-field dividing each flat-field image by its median value, excluding bad pixels (Figure \ref{dark-confrnto}.a).
The bad pixels correction was performed in two steps. In the first step, we used the bad pixel map provided by ESO\footnote{https://www.eso.org/sci/facilities/paranal/instruments/eris/img/master\\ \_bpm\_lamp.fits.gz}, which was obtained using an internal lamp within the instrument. We then created an \enquote{instantaneous} bad pixels map fitting the histogram of pixel values of the entire image with a Gaussian. Pixels with a value higher or lower than 5$\sigma$ from the mean of the Gaussian were defined as bad pixels. By combining the two maps (Figure \ref{dark-confrnto}.c), we generated a final bad pixel map, which slightly differs from the ESO-provided one due to the sensor's evolution over time. This updated map was then applied individually to each science frame, master dark, and master flat to remove the bad pixels.
Finally, the bad pixels were replaced with the linearly interpolated value of the closest neighboring pixels. However, large damaged areas, in the central and lower part of the sensor (Figure \ref{dark-confrnto}.d) cannot be corrected and, therefore, were excluded from the correction and analysis. For this reason, the science target is accurately placed and recorded on the upper part of the sensor, far from these damaged areas.\\

\subsection{Background subtraction}  \label{Background subtraction}
\begin{figure}[h]
    \begin{adjustwidth}{-0\textwidth}{}
        \centering
        \includegraphics[width=0.5\textwidth]{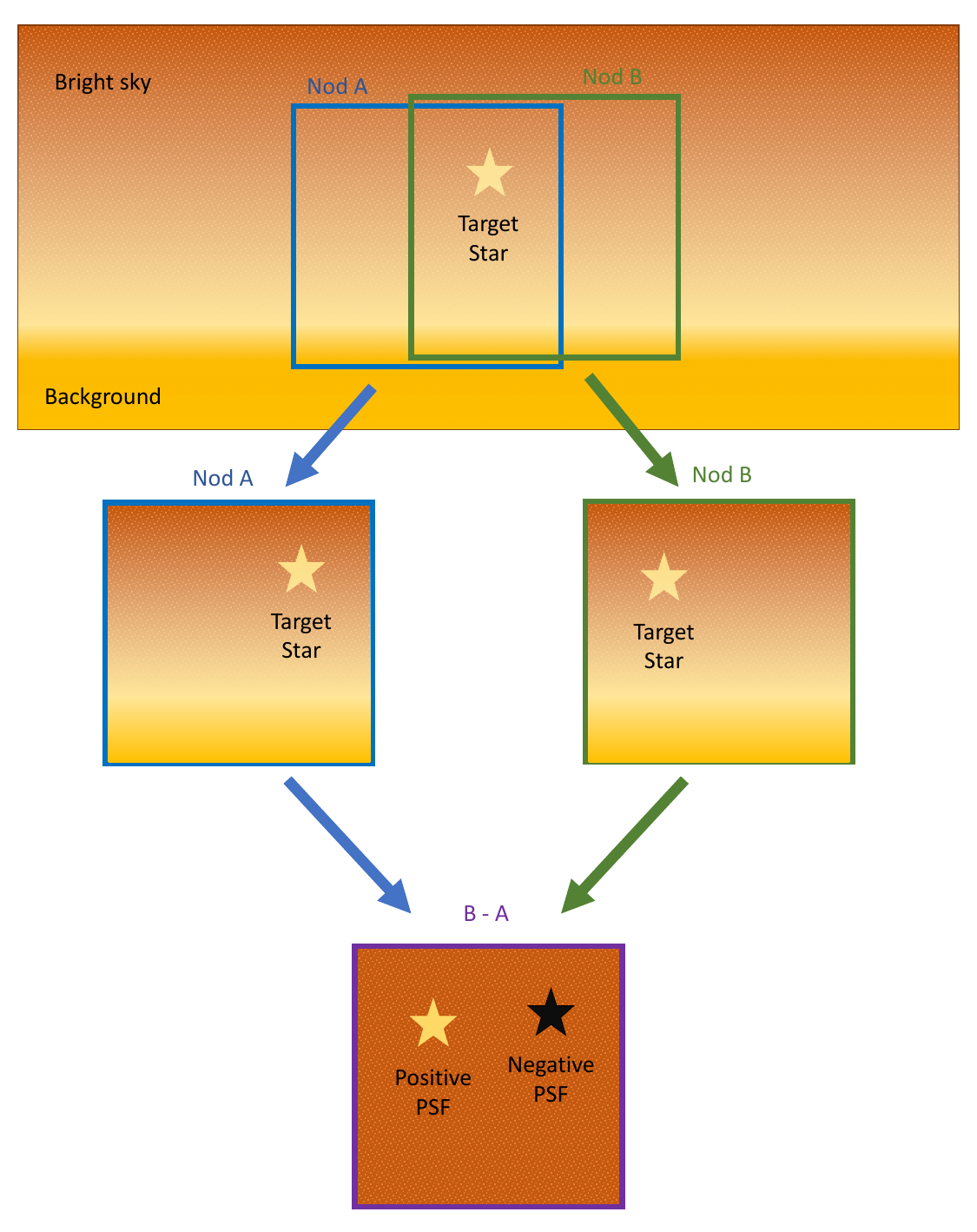}
    \end{adjustwidth}
    \caption{Diagram showing the process of background subtraction by observing two nodding positions in the sky, always having the target star in the field. By subtracting the two exposures, we can suppress the background.}
     \label{differenzanodia}
\end{figure}

Taking advantage of the pupil plane coronagraph, we performed an ABBA nodding sequence with a nodding throw of $\sim9\farcs75$ ($\sim750$ pixels) to keep the star always in the detector and to avoid the superposition of the PSFs during the AB sequence.
The center of the leakage-term is estimated (with sub-pixel accuracy) through a fit of a zero-order bivariate Gaussian.
To analyze the background stability in each frame, we selected two regions of the background and studied the variations in intensity and standard deviation as a function of time. No variations were detected, except for HD 169142, where spatial variations in the background were observed in the final image, along with temporal variations within the data cubes.\\
We created the master background images for each nodding position: when the star was in position B by averaging all the frames in position A and vice versa. We subtracted the master background \enquote{A} from the exposure in which the star was in position A and vice versa, following \citet{2020AJ....159..252W}, \cite{Sutlieff.app} and \cite{cugno.naco}.
To minimize the sky-related background variation, we subtracted the master background generated from the exposure closest in time, as seen in Figure \ref{differenzanodia}.

The median of all frames in each exposure still showed a series of uncorrected vertical and horizontal bands, remaining from previous stages of calibration. This was corrected by masking all the structures related to the PSF in the image and subtracting the median pixel intensities of each column from all pixels in that column. We subtracted the median pixel intensities to account for the non-Gaussian distribution of the measurements caused by the potential residual presence of the PSF.

\subsection{Frame alignment} \label{Alignment}
After subtracting the background from all the frames within an exposure, we aligned the PSF center to the frame's center. To recenter the images, we used the same approach as \cite{Sutlieff.app}, defining the PSF center fitting a bivariate Gaussian. The Gaussian sigma is $3.35 \pm 0.16$ pixels (corresponding to $\sim 0$\farcs$043$) which is comparable to the PSF of the VLT at 4 $\mu$m. The alignment achieved in our analysis reaches an accuracy of approximately one tenth of a pixel. This precision is critical, as the pixel counts generated by the presence of a disk or planetary companion were comparable to the noise level (less than 1$\sigma$) in the single frame. Therefore, subpixel alignment is essential for effectively capturing and summing the photons emitted by the planet or disk, ensuring reliable detection and characterization of these faint signals.\\

\subsection{Frame selection} \label{Frames selection}
Before co-adding all the reduced frames within each exposure, we performed a selection to exclude those frames affected by bad seeing and AO open loops. The median of the Gaussian sigma of the central PSF leakage-term was used in the fit for the centering and it was applied a selection threshold based on the median absolute deviation (MAD). 
The MAD is defined as 

\begin{equation}
    \text{MAD} = \text{median}(| \sigma_i - \tilde{\sigma}) |),
\end{equation}
where $\sigma_i$ is the standard deviation of individual frames and $\tilde{\sigma}$ is the median of all the frames in a given exposure. We excluded all the frames with a width above 5 MAD.

\subsection{Reference differential imaging} \label{Residual PSF subtraction}
Before extracting the two dark holes from each exposure we subtracted the residual of the vAPP PSF by means of a PSF reference. We reconstructed the vAPP PSF reference by taking the median of the frames acquired during the observation night for the stars TW Hya, AS 209, and Elias 2-24 (Figure \ref{PSF BS phot}, bottom-right panel). By taking the median of all the good frames of the cube, any extended disk signal present in the individual frames of the science targets is canceled out.
Additionally, we analyzed all the available vAPP observations in Br-$\alpha$-cont conducted so far, including those taken during the commissioning phase of $\beta$ Lib, which does not show any evidence of circumstellar structures. For each star observed in our study, we used the combination of reference stars that provided the best subtraction of the PSF.
We then subtracted the median vAPP PSF reference from each frame of the science targets via reference differential imaging (RDI). To do this, we first normalized the PSF reference to match the integrated flux of the individual frames.\\

The narrow bandwidth of the Br-$\alpha$-cont filter ensures that spectral slope differences between stars of varying spectral types have little effect on the PSF morphology. Unless the target exhibits strong emission or absorption features in Br-$\alpha$-cont, the shape of the reference PSF remains a reliable match. Additionally, the reference PSFs were normalized to unity before being scaled to match the target star's brightness using an RDI approach, further mitigating any residual spectral mismatch.

\subsection{Dark holes extraction}
\begin{figure*}[h]
    \centering
    \includegraphics[width=\textwidth]{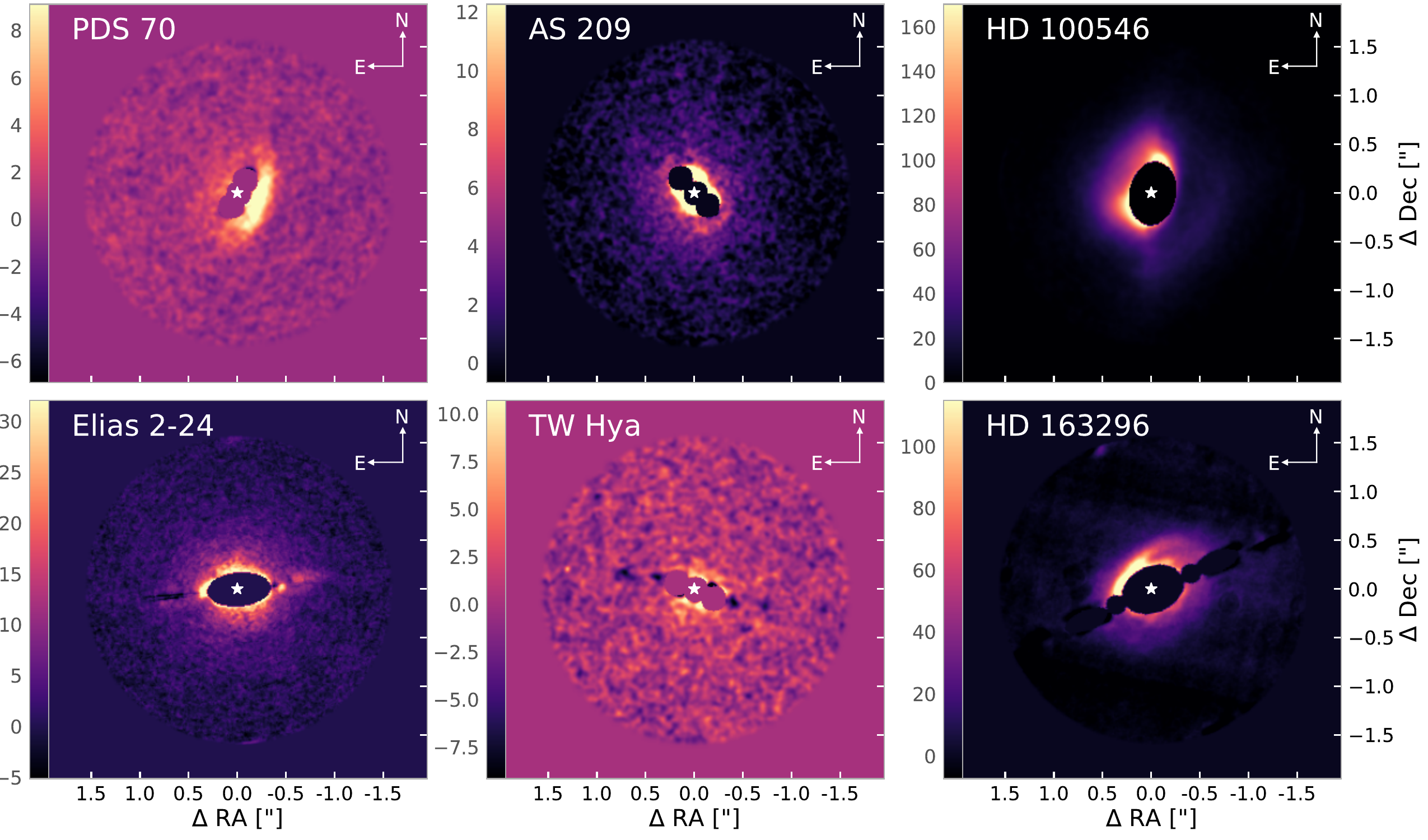}
    \caption{RDI image gallery of observed targets. In the images, the internal portion where residuals of the vAPP PSF were present was masked. In the case of HD 163296, it was also necessary to mask the remaining part of the PSF due to the saturation. Color scale refers to counts.}
    \label{RDI}
\end{figure*}

The next image processing step involves extracting the two dark holes from each datacube, already aligned and PSF-subtracted.
First, we rotated the images by 36$^\circ$, which is the position angle of the vAPP relative to the telescope pupil. This value is derived from the user manual (v114.0\footnote{https://www.eso.org/sci/facilities/paranal/instruments/eris/doc/ERIS\ \\ \_User\_Manual\_v114.0.pdf}) and subsequently validated through tests carried out on our data (Appendix \ref{PSFcarat}). The image was translated to align the center of the upper PSF with the image center. Subsequently, the mask was applied to extract the two dark holes. The mask is a semicircle with a radius of 1\farcs56 (120 pixels) centered on the PSF center. We repeated the same procedure for the lower PSF using the same (rotated) mask. Finally, the PSF-subtracted images were obtained by combining the two extracted dark holes. To be consistent with the next steps of the analysis, we again rotated the images by -36$^{\circ}$. 
A detailed characterization of the vAPP PSF is presented in Appendix \ref{PSFcarat}.
The final PSF-subtracted images (obtained through reference differential imaging) are shown in Figure~\ref{RDI}.
To assess the impact of dark hole merging on the signal from potential planets and disk structures, we performed simulations. The results show no flux loss for planets at angular separations beyond $0\farcs25$ when transitioning between dark holes. Instead, an unexpected flux increase is observed. In contrast, planets within $0\farcs25$ experience a flux decrease. The simulations are presented in Appendix~\ref{fakeplanet-TZ}.

\subsection{Angular differential imaging}
As the data were taken in pupil-stabilized mode, we also performed classical Angular Differential Imaging (cADI) for the targets observed with a field rotation greater than 10$^{\circ}$; that is, TW Hya, AS 209, PDS 70, and Elias 2-24. 
The field rotation is the difference of the position angle (PA): the angular measurement of the instrument field orientation on the sky. The PA is measured in degrees, counterclockwise from north through east, relative to the celestial north pole.
The PSF of the star was reconstructed by computing the median of the non-rotated images. This PSF was subtracted from each frame, after which all frames were derotated and combined using median stacking.
In a second step, we performed ADI with a principal component analysis (PCA) using the Vortex Image Processing package (\texttt{VIP})\footnote{https://vip.readthedocs.io/en/latest/index.html} package \citep{2017AJ....154....7G,2023JOSS....8.4774C}. 
We injected fake planets by adding a PSF from the bright side, upper and lower, into the corresponding dark hole, scaled by the contrast of the planet. The analysis of the residuals suggests that subtracting more than 10 PCs leads to severe over-subtraction of the very inner structures. After several tests, we found that the trade-off between PSF residual removal and over-subtraction is reached with a number of principal components between four and five.

\begin{figure*}[h]
        \centering
        \includegraphics[width=0.77\textwidth]{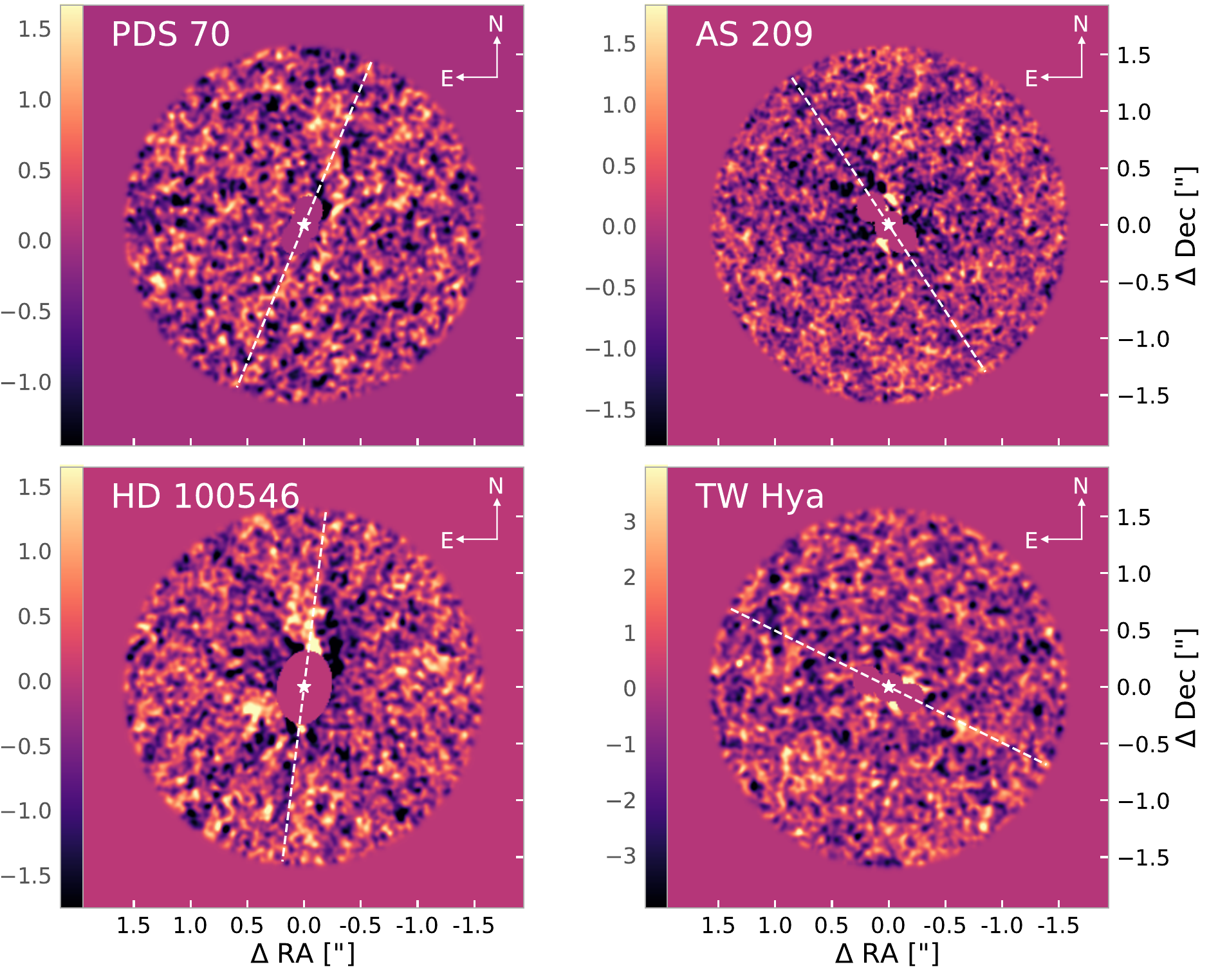}
    \caption{Image gallery of disks made using the classical ADI method. Color scale refers to intensity with arbitrary units. The dashed white line indicates the average position of the junction between the two dark holes during the observation.}
    \label{cADI}
\end{figure*}

\begin{figure*}[h]
        \centering
        \includegraphics[width=0.77\textwidth]{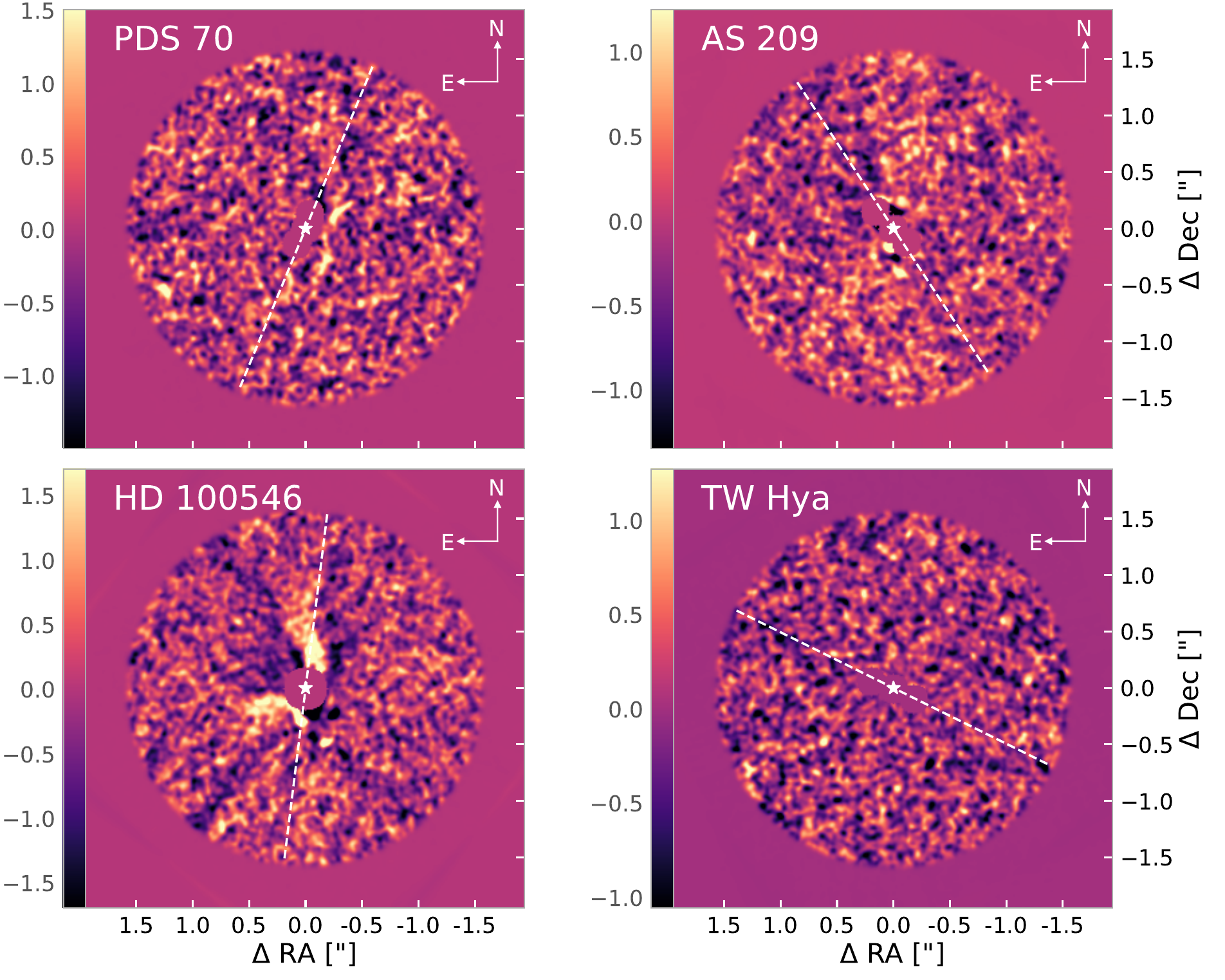}
    \caption{Image gallery of disks made using the ADI-PCA method. Color scale refers to intensity with arbitrary units. The dashed white line indicates the average position of the junction between the two dark holes during the observation.}
    \label{ADI-PCA}
\end{figure*}

The vAPP PSF core is inherently asymmetric because, when merging the two dark holes, residual second-order PSF structures from the bright side remain, contaminating the dark hole. This asymmetry can introduce artifacts that may be misinterpreted as planetary signals or disk structures. To eliminate this issue, in Figures \ref{cADI} (cADI) and \ref{ADI-PCA} (ADI-PCA), we have masked the central region of the PSF, including the stellar position and the two first-order diffraction peaks. This results in a three-lobed mask that prevents contamination from the PSF structure in the final reduced images.\\


\subsection{Contrast curve} \label{contrast_curve_section}
\begin{figure*}[h]
    \centering
    \includegraphics[width=0.9\textwidth]{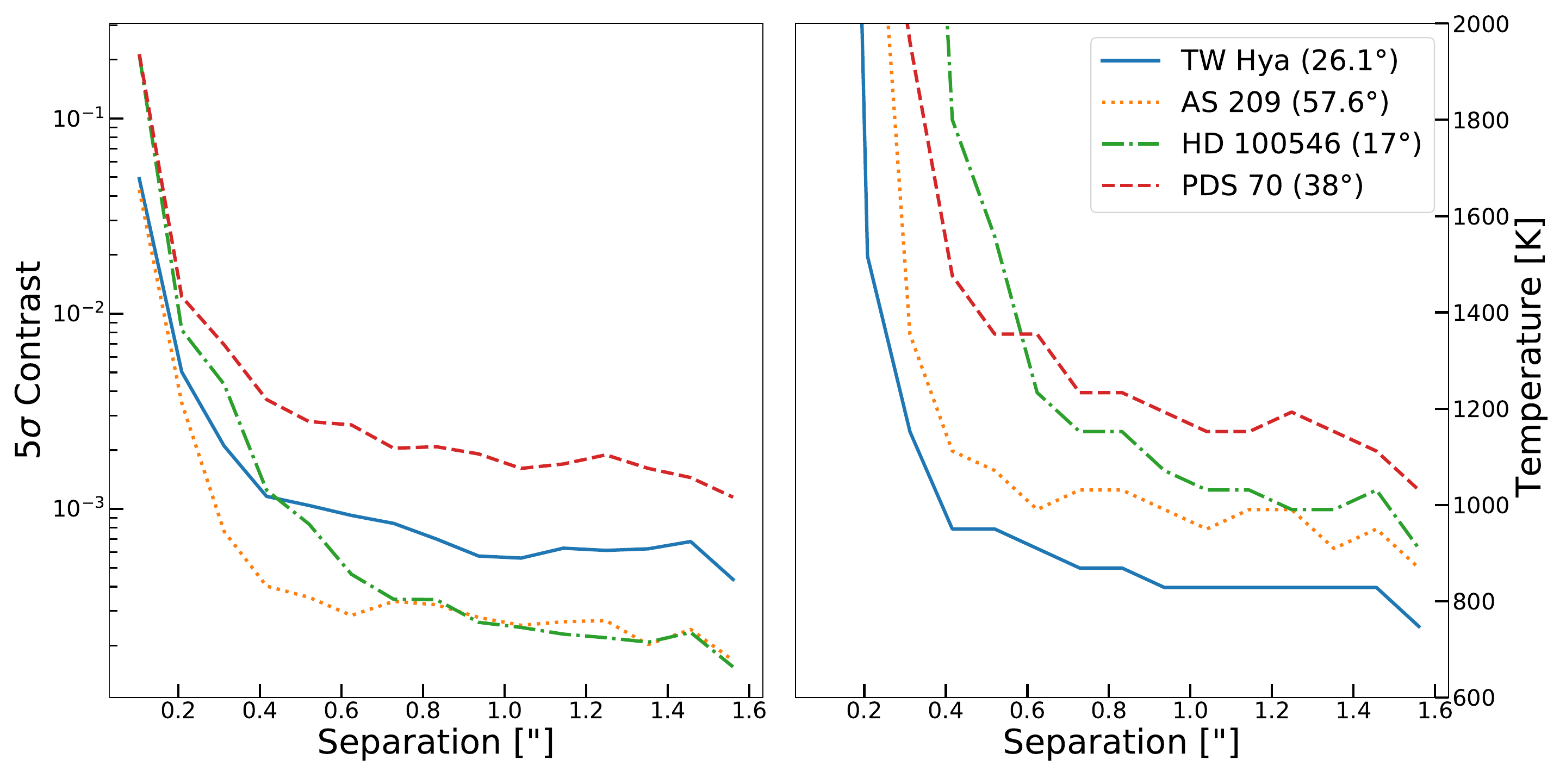}
    \caption{Left: Contrast curves for the various targets. A dependency between contrast and field rotation is visible. PSD 70 deviates from this trend as it is the faintest star in our sample. Right: Planet temperature upper limit curves ($R_p = 2\,R_\mathrm{Jup}$) for the four targets.}

    \label{contrasti}
\end{figure*}

We measured the contrast curves in the ADI-PCA images; that is, for TW Hya, AS 209, PDS 70, and Elias 2-24. To derive the contrast, we computed the standard deviation of the noise within an annulus spanning multiple radial distances around the star. The contrast at a given distance is then determined by the ratio of this noise to the stellar flux, divided by the throughput. The throughput is estimated by injecting a simulated planet with known flux at a certain distance from the star and then recovering the planet's flux after post-processing to determine the flux ratio. \\

To derive the contrast curve, we use the Python package \texttt{AppleFy} \footnote{https://applefy.readthedocs.io/en/latest/index.html} \citep{bonse2023comparing}, which performs the same steps described earlier (fake planet injection, photometry, throughput calculation) using ADI-PCA from \texttt{VIP}. It also accounts for small sample statistics by adopting the t-distribution for the test statistic under the assumption of Gaussian noise. The stellar photometry estimate was done using one of the two bright PSFs from the vAPP. The simulated planets were injected as a rescaled stellar PSF into each frame after the merging of the two dark holes.\\
We employed the dedicated procedures within \texttt{VIP} to compute the signal-to-noise ratio (S/N) and significance. To do this, we applied the small sample statistic of \citet{Mawet2014}, which is the standard correction in the \texttt{VIP} package.
The 5\,$\sigma$ contrast curves for the four stars in our sample are shown in Figure~\ref{contrasti}. Our analysis has shown that the plateau level of the contrast can be explained by the background photon noise contrast level, as presented in the Appendix \ref{PhotonNoise}.\\

Using the previously estimated 5-$\sigma$ contrast curves, we calculate the lowest effective temperature a planet can have to be detectable in the data. This process revisits the method proposed by \citet{cugno.naco}.\\
We fit the best contrast achieved in the Br-$\alpha$-cont filter with the flux corresponding to that contrast, assuming a black body emission. This simple model uses the planet's radius as a free parameter; given the achieved contrasts, we chose to explore a range between 1 and 3 Jupiter radii. 
This procedure requires significantly fewer assumptions than standard atmospheric and/or evolutionary models, and we consider it more appropriate given that we only have a single photometric point in our observations.

\begin{table}[h!]
\centering
\caption{Parameters used for the blackbody of the stars.}
\begin{tabular}{cccccc}
\hline
\textbf{$Target$} & \textbf{$dist. [pc]$} & \textbf{$L' [mag]$} & \textbf{$R_*[R_\odot]$} & \textbf{$T [K]$} & \textbf{$refs$}\\ 
\hline
\hline
AS 209 & 121 & 6.22 & 2.26 & 4250 & 1\\ 
PDS 70 & 112 & 7.93 & 1.26 & 3972 & 2\\ 
HD 100546 & 108 & 5.87 & 1.50 & 9800 & 3,4 \\ 
TW Hya & 60 & 6.97 & 1.10 & 4000 & 5\\ 
\hline
\end{tabular}
\tablebib{1.\citet{2018A&A...610A..24F}, 2.\citet{KepplerPSD702018}, 3.\citet{citazione14}, 4.\citet{2015MNRAS.453..976F}, 5.\citet{2007ApJ...660.1556R}.}
\label{tab:star_parameters}
\end{table}

We used a simple double blackbody model. The first blackbody represents the star and the second is the planetary spectrum. The first is characterized by its temperature and rescaled by the stellar surface area. The temperature and radius of each star are listed in Table \ref{tab:star_parameters}. For the second, we created a grid with temperatures ranging from 100K to 6000K in 100K increments, and planetary radii of 1, 2, and 3 $R_J$. For each planetary radius, we calculated the luminosity ratio between the star and the planet. Assuming that both bodies are at the same distance, this ratio approximates the flux ratio, yielding the contrast.\\
The luminosity, $L(\lambda)$, of each body was computed with:
\begin{equation}
L(\lambda) = 4 \pi R^2  B(T, \lambda),
\end{equation}
where $R$ is the radius of the body, and $B(T, \lambda)$ is the spectral radiance defined by the Planck law.
After computing the contrast in the Br-$\alpha$-cont filter, we compared it with the observed contrast curves. This allowed us to estimate the temperature that best matches the measured contrast at each separation from the star. We generated a curve of minimum detectable temperature versus angular separation for each planetary radius. In our analysis, we excluded local disk extinction.
The results are shown in Figure~\ref{contrasti} for the four stars in our sample for a fixed planet radius (2 R$_J$), while Figure~\ref{TEMP_LIM} in Appendix \ref{Temperaturelimitplot} shows the different values of the planet radius.

\section{Results}
In this section, we discuss the characterization of the disk signal revealed via direct imaging at 4$\mu$m obtained via RDI. We further discuss the potential point-source candidates detected via ADI.\\
In this section, we will discuss the results as observed in the data. A more detailed analysis and comparison with the literature will be presented in Section \ref{fine}.\\
While the observation was optimized for detecting extended signals, such as the disk, we also applied ADI for targets with a field rotation greater than $10^\circ$. The aim was not to achieve very high contrast but rather to test the performance of the vAPP and the pipeline's capabilities. Moreover, this approach will be important for next-generation instruments operating with vAPP coronagraphs, such as METIS at ELT \citep{Brandl2021}.\\

\subsection{Reference differential imaging}
\begin{figure*}[h]
    \centering
    \includegraphics[width=0.9\textwidth]{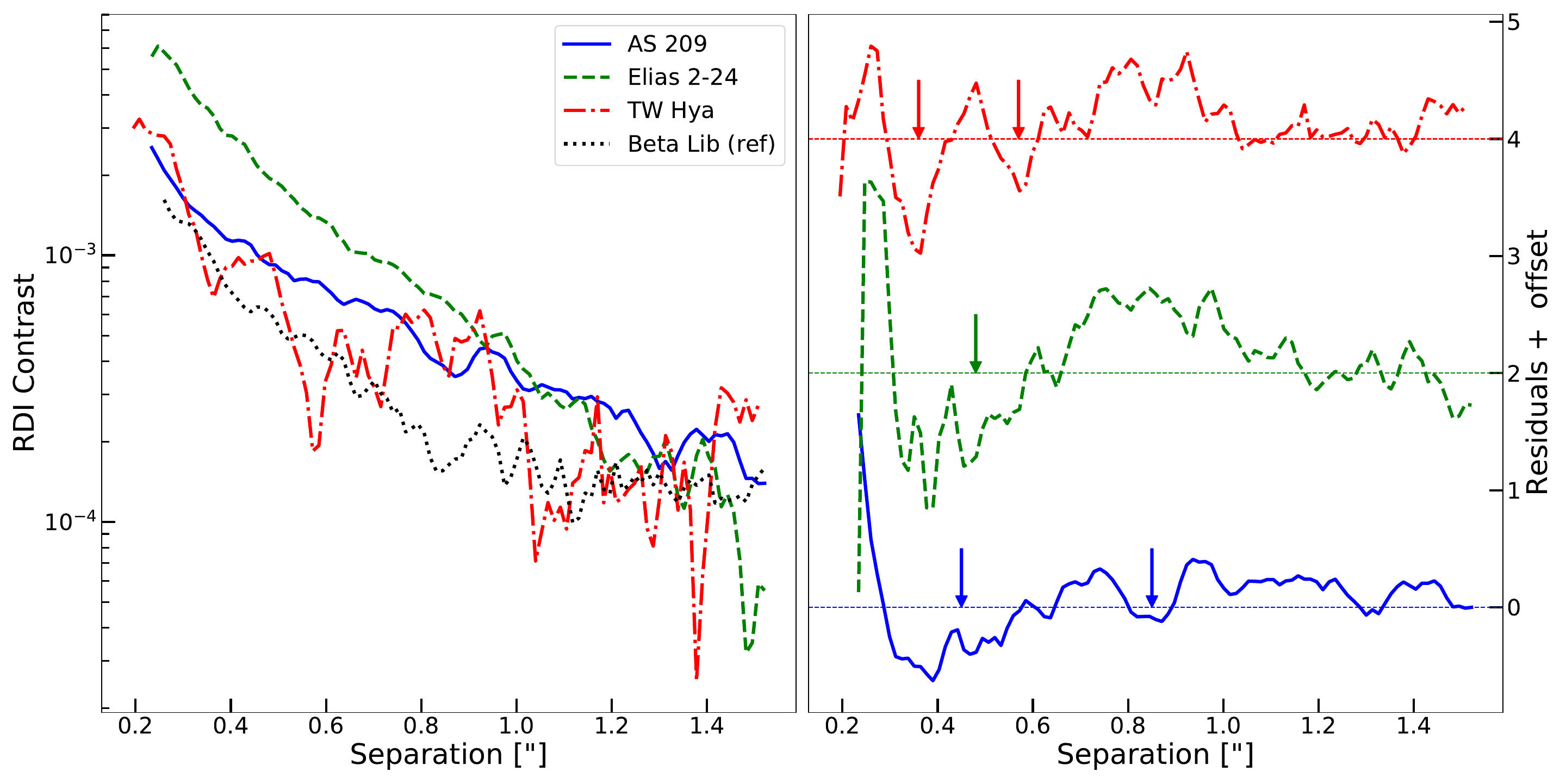}
    \caption{Left: Radial profiles after RDI of TW Hya, AS 209, Elias 2-24, and the reference star $\beta$ Lib in logarithmic scale. The profiles were computed from RDI data. Separation is expressed in arcseconds, and the flux is normalized to the stellar peak intensity. Right: Residuals of the radial profiles after exponential fitting, with an incremental offset of 2 applied to each curve to avoid overlap. A dashed line indicates the new zero level of the residuals and the arrows indicate the observed gaps.}
    \label{radial-profile}
\end{figure*}

In all the systems, spatially resolved extended disk signal is detected in the reference differential imaging (RDI) images (Figure \ref{RDI}). In particular, we resolved for the first time mid-infrared emission from the disks around AS 209 and Elias 2-24. The emission in these two systems is axisymmetric and extends out to 55-60 AU. The radial profile, for the most face-on disks (i< $40^\circ$), was extracted only from regions of the field unaffected by PSF residuals (present along the vAPP position angle). The selected azimuthal ranges are $60^\circ$ to $146^\circ$ and $242^\circ$ to $333^\circ$ for TW Hya, $36^\circ$ to $121^\circ$ and $225^\circ$ to $307^\circ$ for Elias 2-24, and $81^\circ$ to $182^\circ$ and $272^\circ$ to $360^\circ$ for AS 209. Figure \ref{radial-profile} shows the radial profiles of these three systems.
As a reference, the radial profile of the star $\beta$ Lib, observed during the commissioning phase and known to lack any circumstellar disk, was also analyzed. This comparison highlights how the signals from AS 209 and Elias 2-24 consistently exceed the reference level. Beyond approximately 1" all profiles converge, either indicating the end of the disks or the noise floor level.

\begin{figure}[h!]
    \centering
    \includegraphics[width=0.4\textwidth]{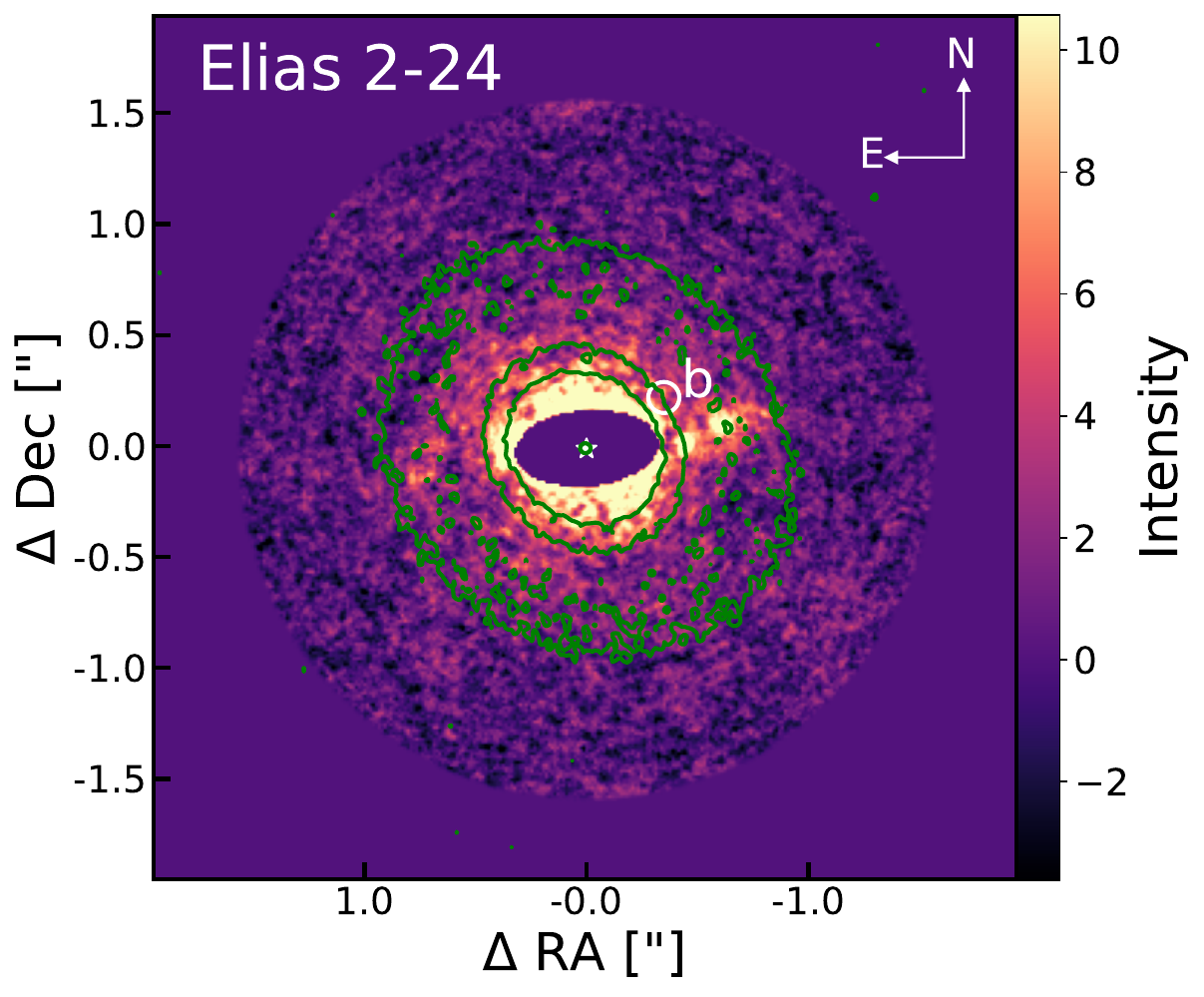}
    \caption{Result of RDI Elias 2-24. Significant emission is noticeable in the inner regions of the disk compared to the gap observed by ALMA (green contours). The position of  \enquote{b} indicated in the image refers to the candidate of \citet{Jorquera2021}. The two blobs visible to the west of the star are remnants of the PSF that were not correctly subtracted during the RDI.}
    \label{Elias 2-24 stacking}
\end{figure}

Extended emission around Elias 2-24 is detected around the star, reaching out to 0\farcs45 from the central mask. This emission matches the orientation of the disk observed with ALMA and corresponds to the first detection of scattered light in the near-infrared, as shown in Figure \ref{Elias 2-24 stacking}. The overlap of the millimeter emission contours from ALMA \citep{2018ApJ...869L..41A} shows that the 4 $\mu$m emission is concentrated within the gap in the millimeter dust. Furthermore, the emission is aligned with the ALMA observations.

By analyzing the radial profile of AS 209, two points where the slope changes can be observed, one at approximately 0\farcs4 (50 au) from the central star and another at around 0\farcs8 (100 au) from the star, in similar way as \citet{2018ApJ...863...44A}in J and H bands. By subtracting an exponential fit from the radial profile and analyzing the residuals, two pronounced gap become evident. These region roughly coincide with the two main gaps observed by ALMA \citep{2018ApJ...869L..41A}. 

In contrast, TW Hya does not show an extended disk in figure \ref{RDI} but only a weak emission within 0\farcs4 (24 au), consistent with the inner and brightest part of the disk. Analyzing the radial profile of TW Hya, the residuals reveal the faint presence of a gap at $\sim0\farcs37$, consistent with the gap observed by \citet{2017ApJ...837..132V}. A second gap is marginally detected at $\sim0\farcs58$, close to the one observed with ALMA by \citet{Andrews2016}.

\begin{figure}[h!]
  \centering
    \includegraphics[width=0.4\textwidth]{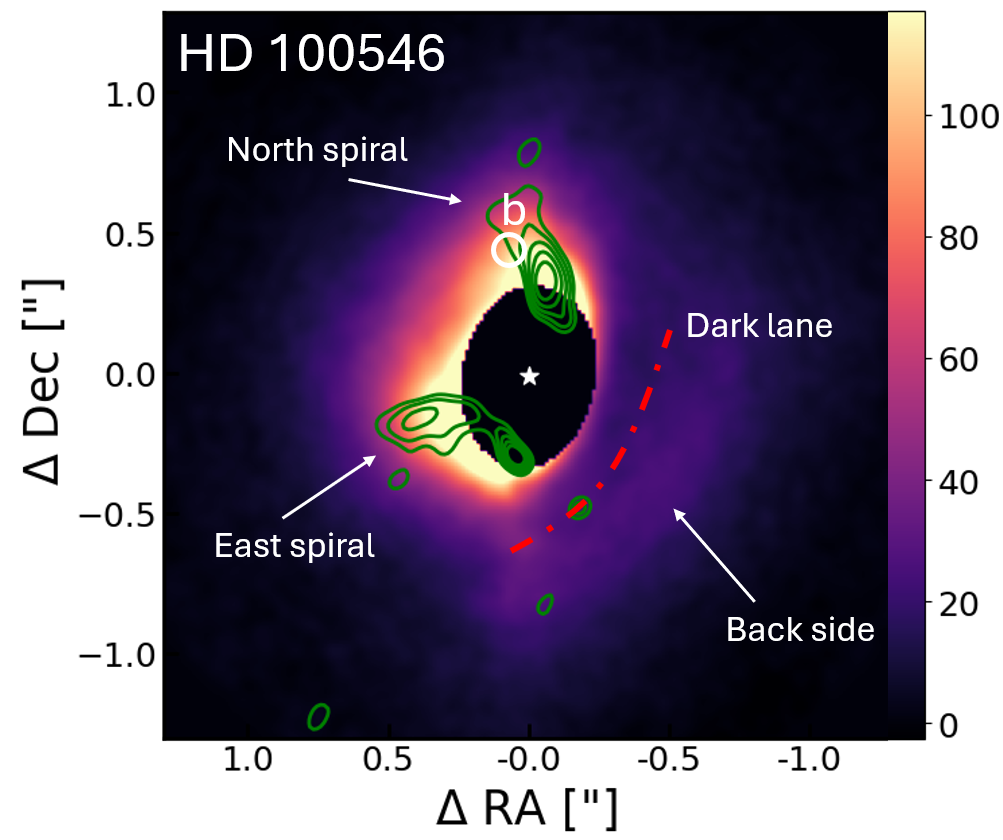}
  \caption{Comparison of structures visible around HD 100546 with RDI and ADI-PCA techniques. The position of  \enquote{b} indicated in the image refers to the candidate of \citet{2015ApJ...807...64Q}. The green contours represent the signal observed in the ADI-PCA.}
  \label{HD100546confronto}
\end{figure}

In the case of HD 100546, the inner part of the image is heavily saturated. Nevertheless, the vAPP image reveals the flared disk structure, showing the dark lane on the near side and the back side of the disk in the west-southwest quadrant. Additionally, faint traces of the two spiral arms are recovered in the northeast region, consistent with the flared geometry of the disk (see Figure \ref{HD100546confronto}).

\begin{figure*}[h]
    \centering
    \includegraphics[width=0.9\textwidth]{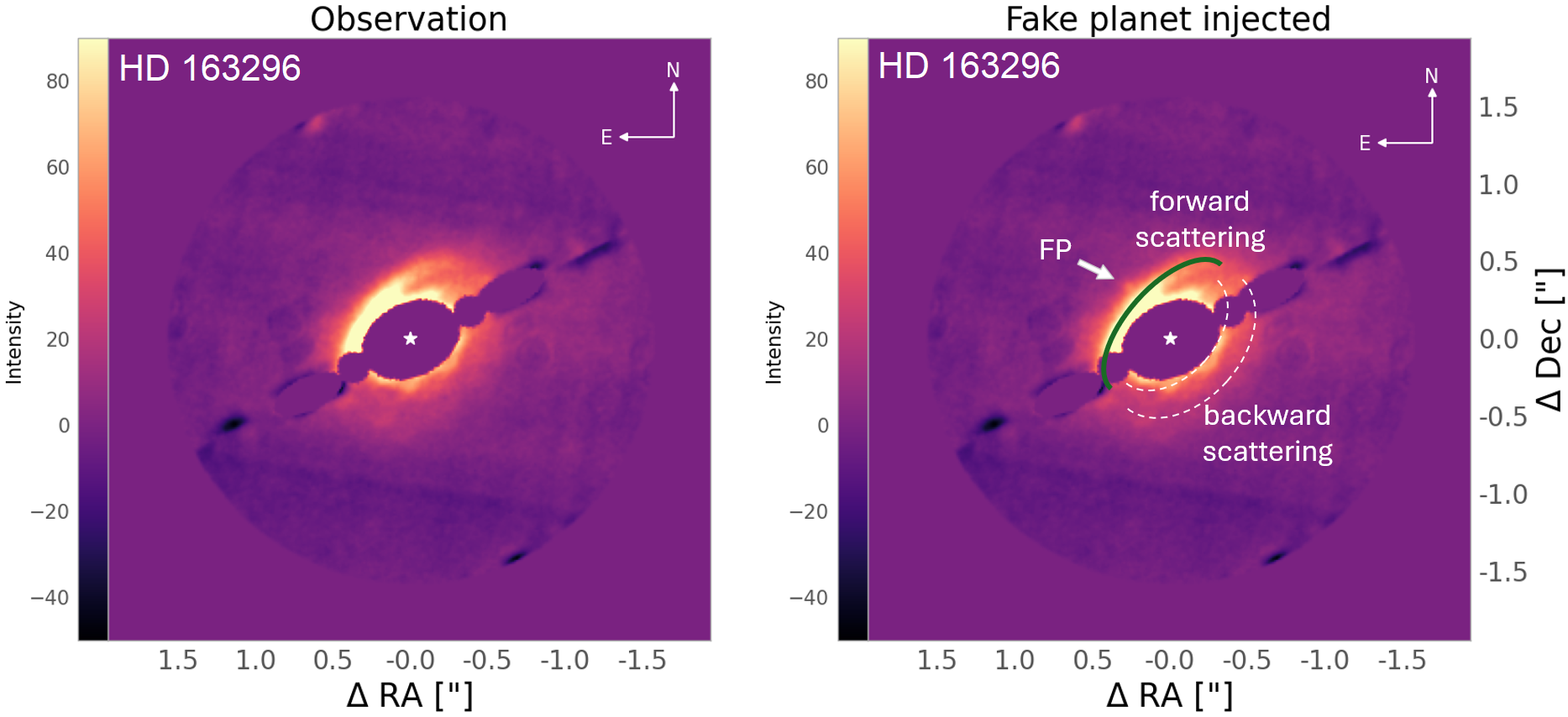}
    \caption{a) RDI of HD 163296. b) RDI of HD 163296 with the simulated planet injected in the correct position, indicated by an arrow. In both images, the object located to the northeast, at the edge of the field, is a remnant of the external PSF from the APP.}
    \label{HD 163296-fake}
\end{figure*}

In the case of HD 163296 an inner dust ring within 0\farcs5 is resolved. The elliptical fit of the ring’s brightness peaks provide a semimajor axis of 0\farcs67 $\pm$ 0\farcs07, a semiminor axis of 0\farcs42 $\pm$ 0\farcs04, and a position angle of $128^\circ \pm 4^\circ$. Backward scattering is observed from the far side of the disk, in addition to forward scattering on the near side. No point-like emission is detected at the position of the candidate planet suggested by \citet{2018MNRAS.479.1505G}. However, injecting a simulated planet matching the claimed signal confirms it would have been recoverable under the observed conditions, as is shown in Figure \ref{HD 163296-fake}.

The disk of PDS 70 is clearly detected, showing the infrared emission in total intensity at 4$\mu$m. Faint emission is visible in the north-northeast quadrant, likely due to backward scattering. Additional infrared emission is detected inside the dust cavity. This is further discussed in a forthcoming paper.

\subsection{Angular differential imaging}
The flux contrasts and the temperatures upper limits shown in Figure~\ref{contrasti} summarize the detection limits reached by our survey in terms of planet detectability. Overall we reached a contrast of nearly 10$^{-3}$ at separation of $\gtrsim 0\farcs3$ and down to $3 \times 10^{-4}$ in the case of AS 209 and HD 100546. These values are comparable to the median value of the contrast curves presented in \citet{cugno.naco} and obtained with NACO with the L' filter. This flux contrast corresponds to a planet temperature upper limit of $\sim$ 800-1000\,K with the most stringent upper limit reached for TW Hya.
A close inspection of the individual ADI-PCA images reveals some candidate structures that merit further investigation.

\textbf{\begin{figure}[h]
    \centering
    \includegraphics[width=0.4\textwidth]{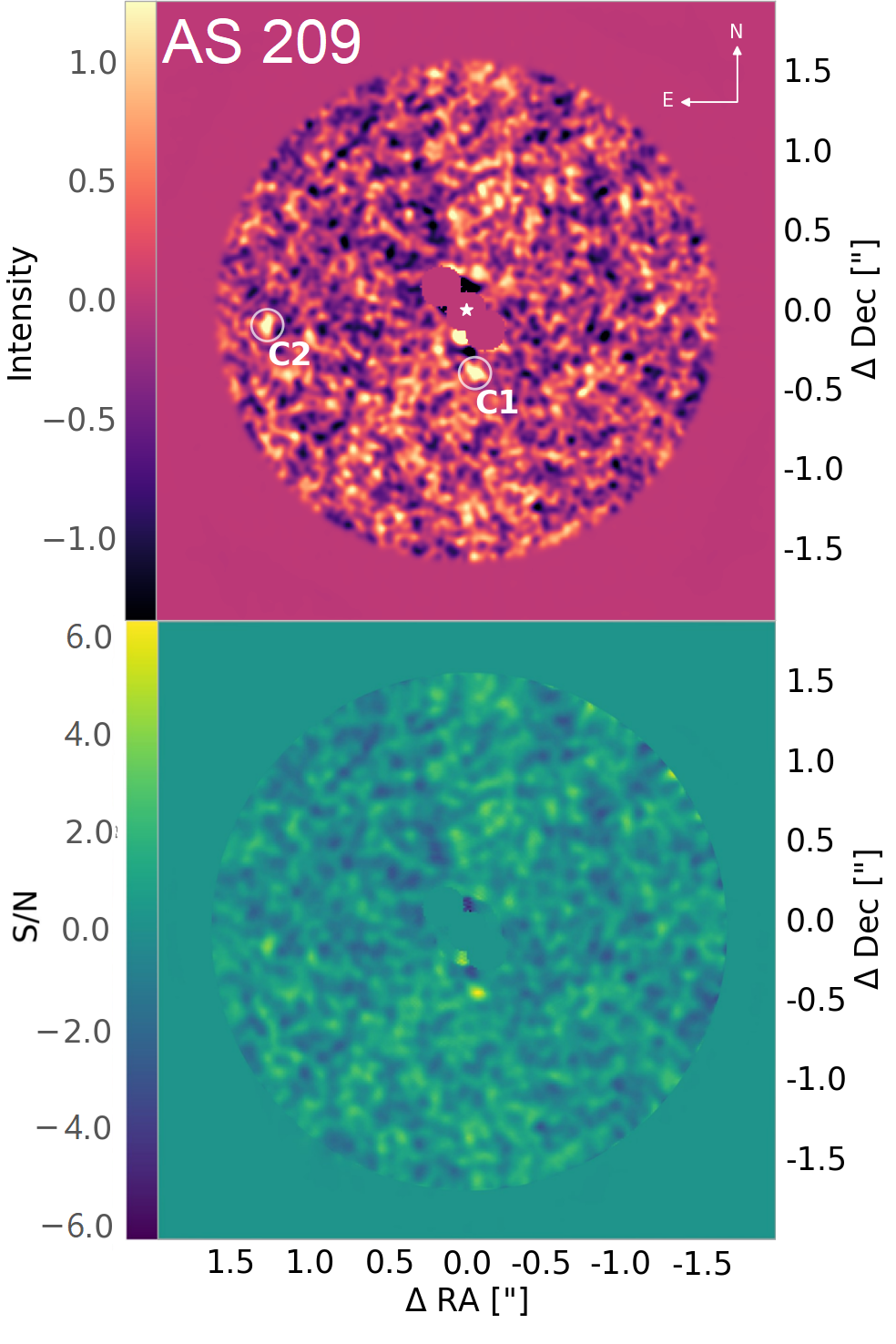}
   \caption{Upper panel: Result of ADI-PCA for AS 209 (three subtracted PCs). Lower panel: S/N map observations obtained with ADI-PCA (two subtracted PCs). }
    \label{AS209-planetPCA}
\end{figure}
}

In our analysis of the AS 209 system, two point-source candidates were identified through S/N map inspection, Figure \ref{AS209-planetPCA}. The two candidates survive both cADI and ADI-PCA analysis, indicating that they are not artifacts related to the analysis technique. These features referred to as C1 and C2, were detected at the same separations from the central star as the gaps observed bt ALMA in mm continuum \citep{2018A&A...610A..24F}. The contrasts and other key properties of these candidates are summarized below.
C1 and C2, located at $0\arcsec.40 \pm 0\arcsec.02$ and $1\arcsec.24 \pm 0\arcsec.02$, respectively, were estimated by injecting artificial planets at the same angular distances from the star. This analysis was performed using the cADI technique to reduce computational time. 
The particular, the strongest candidate, C1 has a position angle of $-75 \pm 3$ degrees and the S/N of this detection is 5.7, corresponding to a significance level of $4.6\sigma$. The measured contrast relative to the stellar flux is $(3.5 \pm 0.2) \times 10^{-4}$. Photometric measurements were used to determine when the injected planets matched the observed counts. 

For HD 100546, the cADI and ADI-PCA techniques recovered two spiral structures located north and east of the star. The two point sources identified by \citet{2015ApJ...807...64Q} were not recovered. At the position of HD 100546 b, only the northern spiral is visible, while the position of HD 100546 c is obscured by the extended PSF mask (see Figure \ref{HD100546confronto}). 
At smaller separations, however, the presence of the spiral structures in the disk limits the achievable contrast.

In the case of TW Hya, both cADI and PCA reductions show a smooth background without structures. The contrast curve reaches a plateau at $6 \times 10^{-4}$, slightly higher than for AS 209 and HD 100546, likely due to the reduced field rotation during these observations.

Even though the detailed study of PDS 70 will be presented in a separate paper, we briefly discuss the relevant findings. PDS 70 b is not detected in our observations, likely due to its position behind the first-order stellar PSF residual, specifically along the apparent position angle.
Regarding planet PDS 70 c, we observe a spiral-like feature extending from the disk toward the position of c. While the structure is located in a similar position to that observed by \citet{Wang2020}, it is morphologically different. This feature suggests possible dynamic processes occurring in the system, which could be associated with the candidate's position and may offer clues about the interaction between the disk and the planetary companion.

\section{Analysis} \label{fine}
In this section, in the first section, we discuss our results comparing them to previous multi-wavelength observations, and in the second part, we propose a new observational method.\\

\subsection{Protoplanetary disks}
Analyzing the case of TW Hya, Elias 2-24, and HD 163296 we observe that none of these objects has revealed low-mass planetary companions associated with the presence of gaps in the PPD. This implies that the minimum mass required to carve gaps in the disks is below the detection threshold, that planets are much colder than 800-1000\,K (for 2 $R_J$), or that the extinction caused by dust still present in the gaps is  $A_V$ > 20 mag (Assuming a mean case of a $2 \, R_\mathrm{J}$ planet with an effective temperature of $2000 \, \mathrm{K}$, orbiting a $1.2 \, R_\odot$ star with a temperature of $6800 \, \mathrm{K}$, and a contrast of $3 \times 10^{-4}$).
These results are consistent with the theoretical predictions by \citet{Ruzza2024, Ruzza2025}, which suggest that the gaps in these systems can also be carved by sub-Jovian mass planets. Such low-mass planets would be cooler and have smaller radii than those detectable in this study.\\

TW Hya is the closest star to the Solar System studied in this work, located approximately 60 pc away. Given the very low inclination of the disk and the presence of multiple gaps observed in the millimeter continuum \citep[e.g.,][]{2021A&A...648A..33M}, this is the system that could most likely present evidence for the presence of planets in formation. Our observations did not reveal any emission from the disk. Analyses through the contrast curve yield a minimum upper limit for the 5$\sigma$ contrast of a planet of $\sim6\times10^{-4}$ corresponding of a limit of $\sim800$ K for a 3 $R_J$ planet. 
In the gap at 25 au, we reach a temperature limit of 990 K, at 40 au a limit of 885 K, at 48 au 870 K, and at 88 au 790 K. These limits are estimated for a planet with a radius of 2 $R_J$.
This value is consistent with literature values, as seen in the study by \citet{2017AJ....154...73R} (observations performed at the same wavelength as in our study), which did not find any candidate planets in the system.\\

Regarding Elias 2-24, the RDI resolved for the first time an extended emission at 4 $\mu$m. Figure \ref{Elias 2-24 stacking} shows that the infrared emission corresponds to the inner disk resolved by ALMA \citep{2018MNRAS.475.5296D}.
This is consistent with a $r^{-2}$ dependence of stellar radiation illuminating the disk or by an actual lack of micrometric dust in the outer part of the disk.

HD 163296 was the brightest star observed in this campaign, hosting one of the brightest disks, which was detected in every individual frame prior to combining and RDI processing. \citet{2018MNRAS.479.1505G} reported a planet candidate with an estimated mass of around 5 $M_J$, located just beyond the ring resolved in the infrared wavelengths. This planet candidate was also not confirmed in L band by \citet{cugno.naco}. Moreover, our observations did not identify any point-like source at that position. To verify our detection limit, we injected a fake planet with the same contrast as the previously reported candidate, and it was successfully recovered. 
In the literature, two additional planet candidates have been proposed based on dynamical gas analysis, showing two distinct kinks \citep{2020ApJ...890L...9P}. The first candidate, northeast of the star and more than 1$\farcs$5 from HD 163296, lies outside the dark hole, and the limited field rotation prevented us from applying cADI and ADI-PCA techniques to this position. The second candidate, located west of the star, was not detected, implying it has a contrast lower than $\Delta L \approx 11$ mag, which is consistent with our detection limits.\\

HD 100546 hosts a complex PPD with features such as a cavity, dust rings, and spirals. Previous studies, including \citet[e.g.,][]{2015ApJ...807...64Q,cugno.naco} with NACO in the L' band, \citet[e.g.,][]{2016A&A...588A...8G,2018A&A...619A.160S} with SPHERE in the J,H and K bands, and \citet{Follette2017} with GPI and MagAO identified two spiral arms in this system. Our ADI-PCA analysis confirms that these spiral arms are consistent in both shape and orientation with those previously observed. Furthermore, no evidence of point sources was found at the position noted by \citet{2015ApJ...807...64Q}, according to the non-detection by \citet{cugno.naco}. The improved performance of the ERIS adaptive optics allowed us to distinguish extended objects from point sources. The performance of ERIS for NGS targets with R magnitudes between 8 and 10 achieves Strehl ratios above 0.8 in the Ks band \citep{Riccardi2022}. In comparison, NACO in the same magnitude range and band achieved Strehl ratios of 0.6 for NGS \citep{Kasper2004}.\\

The last system of our sample is AS 209's disk, which hosts one of the most structured PPD resolved with ALMA in our sample. The dimensions and position of the multiple gaps have been interpreted as the potential presence of multiple planetary companions sculpting the structure \citep{dischi1}. The position of C1 is also highly contaminated by these artifacts and disk residuals, making it difficult to determine the existence of this point source.
As for candidate C2, in the observations by \citet{cugno.naco}, it is situated in a much \enquote{cleaner} region. A preliminary visual inspection, of the NACO image, makes it challenging to identify a source in that position, a structure with a very low S/N is barely visible slightly north of the position observed for C2. It is highly likely that this candidate is a noise feature rather than a real signal.\\
Comparing the positions of C1 with the ALMA observations presented in the work by \citet[][]{2018A&A...610A..24F, dischi1}, we observe that C1 falls precisely within the 61 au gap. Numerous studies in the literature, utilizing SPH and fluid dynamics simulations \citep[e.g.,][]{2018A&A...610A..24F,Favre2019}, hypothesize the presence of a planet in this gap. This structure was originally associated with the existence of a second single planet in the 100 au gap. However, a single planet may be capable of carving multiple gaps, potentially explaining why only one candidate has been identified. 
The observations analyzed by \citet{Wallack2024} with Keck/NIRC2 achieve contrasts on the order of $10^{-5}$ and do not reveal any structures at the position of C2. Additionally, no candidate appears to be present at the position of C1. However, as is highlighted in their study, numerous spurious signals are detected in the inner region of the system, caused by residual disk features.
Further investigation of the candidate will be essential to confirm its physical nature.\\

Observations of a kink in the velocity maps of the $^{12}$CO J=2-1 and $^{13}$CO J=2-1 lines suggest the potential presence of a planetary-mass companion at $\sim200$ au \citep{2022ApJ...934L..20B}. Further analysis of the same kink indicates that the planet may reside in a more inner region, at a distance of less than $100$ au \citep{2023A&A...672A.125F}. We have a non-detection for the 100 au ($\sim0\farcs8$) planet, and the upper limit corresponds to a temperature of $\sim 850$ K for a $2 R_J$ object. Using the \texttt{speaces}\footnote{https://species.readthedocs.io/en/latest/} Python package \citep{Stolker2020}, we estimate an upper mass limit of $\sim 8 M_J$. The candidate at 200 au lies outside the dark hole, and assuming a flat contrast curve, we derive an upper mass limit of $\sim 6 M_J$.\\

\subsection{Improved observation technique}
Based on the promising results obtained with RDI, we propose an improved approach for using the APP-like coronagraph specifically designed for studies of circumstellar disk structure rather than planet detection. Observations should be conducted in pupil-tracking mode, allowing the vAPP mask to remain aligned with the telescope spiders throughout the sequence.  A nodding sequence (ABBA) should be implemented to effectively subtract the background. Additionally, it is crucial to save the data in cube mode to accurately characterize the high-frequency temporal background variations. Reference stars should be used for PSF characterization, ensuring they are observed under similar seeing conditions and AO performance as the target. 
The observation sequence should be divided into two stages: the first with the north aligned at the top of the detector, and the second with the instrument rotated by 90° relative to the sky.
This approach enables a more precise reconstruction of the dark hole region by reducing PSF leakage from the bright side, ultimately producing a fully cleaned and refined image suitable for detailed disk studies.

\section{Conclusion}
This study highlights the performance of the vAPP in resolving the morphology of PPDs in total intensity at 4~$\mu$m. Using RDI, the vAPP successfully detected extended structures such as rings, spirals, and cavities. Notably, it enabled the first detection of scattered light emission at 4~$\mu$m from the disks around AS~209 and Elias 2-24. Two potential gaps have been identified in the radial profile of TW Hya at approximately $22\,\mathrm{au}$ and $35\,\mathrm{au}$, as well as two gaps in the radial profile of AS 209 at roughly $50\,\mathrm{au}$ and $100\,\mathrm{au}$.
The RDI also revealed faint structures within the cavity of PDS 70, consistent with ongoing planet formation processes.

In the AS 209 system, two point-source candidates were detected at separations of $0\arcsec.40$ and $1\arcsec.24$, coincident with the ALMA mm continuum gap at 61 au from the star. The innermost candidate, C1, shows the highest S/N of 5.7 but is located near the inner edge of the system's disk. This proximity could explain the point source as a residual artifact from the disk subtraction process.\\

The results, obtained through ADI, demonstrate a contrast limit of $3 \times 10^{-4}$ at 1\arcsec, achieved using a narrow-band filter and modest field rotations of $58^\circ$ and $17^\circ$, as obtained from the contrast curve of AS 209 and HD 100546. Longer integration times and larger field rotations in future observations could further improve this contrast. Dedicated follow-up with instruments like the ERIS annular groove phase mask coronagraph and enhanced techniques will be essential to confirm these findings and advance our understanding of planet formation in PPDs.
These results confirm the ERIS APP's suitability to observe PPDs, providing valuable constraints on disk morphology and the presence of low-mass stellar companion or giant planet. 

\begin{acknowledgements}
We gratefully acknowledge support from the  \enquote{Programma di Ricerca Fondamentale INAF 2022 and 2023} of the Italian National Institute of Astrophysics (INAF GO/GTO grant 2022  \enquote{ERIS \& SHARK GTO data exploitation} and INAF Large Grant 2023  \enquote{NextSTEPS}). 
We acknowledge that the results are based on public data released from the ERIS commissioning observations at the VLT Yepun (UT4) telescope under Programmes ID 60.A-9917(F).
We also acknowledge the valuable contributions of David Doelman, Ben Sutlieff, Valentin Christiaens, Felix Dannert, Markus Bonse, and A.L..
V.R. acknowledges the support of the European Union’s Horizon 2020 research and innovation program and the European Research Council via the ERC Synergy Grant ``ECOGAL'' (project ID 855130). 
A.G. acknowledges financial contribution from PRIN-MUR 2022YP5ACE.
LP and CC acknowledge financial support under the National Recovery and Resilience Plan (NRRP), Mission 4, Component 2, Investment 1.1, Call for tender No. 104 published on 2.2.2022 by the Italian Ministry of University and Research (MUR), funded by the European Union – NextGenerationEU– Project Title 2022JC2Y93 ChemicalOrigins: linking the fossil composition of the Solar System with the chemistry of protoplanetary disks – CUP J53D23001600006 - Grant Assignment Decree No. 962 adopted on 30.06.2023 by the Italian Ministry of Ministry of University and Research (MUR).
LP and CC also acknowledge the EC H2020 project “Astro-Chemical Origins” (ACO, No 811312), the PRIN-MUR 2020 BEYOND-2p (Astrochemistry beyond the second period elements, Prot. 2020AFB3FX), the project ASI-Astrobiologia 2023 MIGLIORA (Modeling Chemical Complexity, F83C23000800005), the INAF-GO 2023 fundings PROTO-SKA (Exploiting ALMA data to study planet forming disks: preparing the advent of SKA, C13C23000770005), and the INAF-Minigrant 2022 “Chemical Origins” (P.I.: L. Podio).
A.Z. acknowledges support from ANID -- Millennium Science Initiative Program -- Center Code NCN2024\_001 and Fondecyt Regular grant number 1250249.
We are grateful to the anonymous referee for his/her valuable report and suggestions.
\end{acknowledgements}
\bibliography{Maioetal_2025.bib}
\bibliographystyle{aa}

\newpage

\begin{appendices}
\renewcommand{\thefigure}{\thesection.\arabic{figure}}

\section{Pipeline flowchart} \label{pipelinescheme}
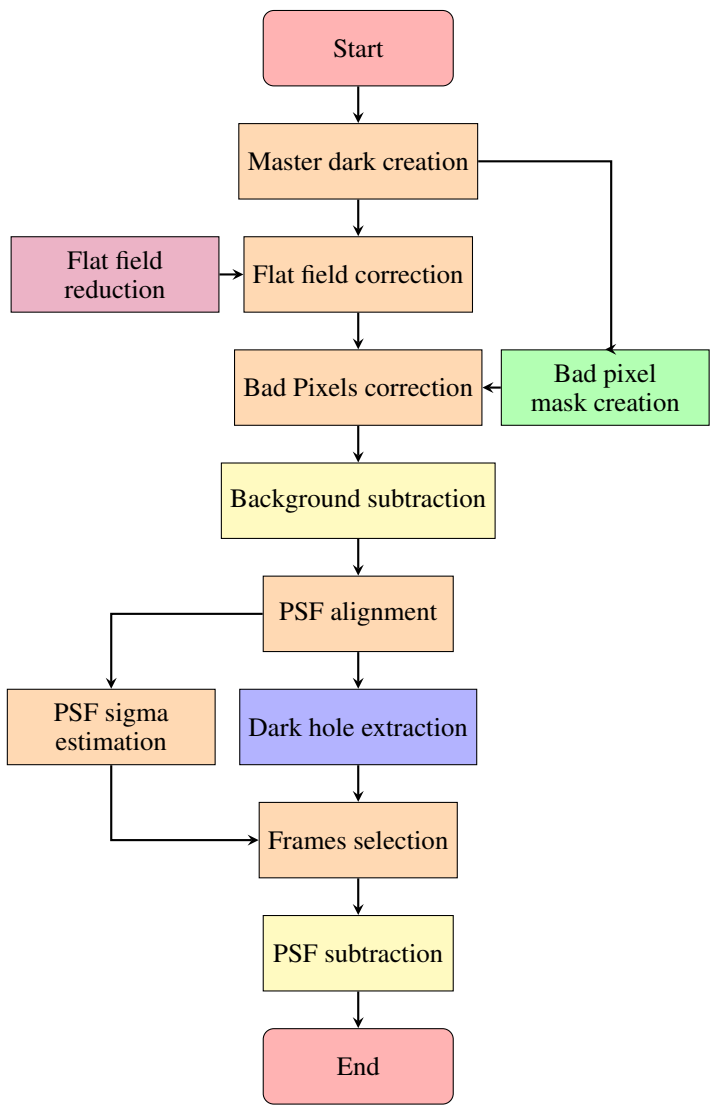
\begin{figure}[h]
    \centering
    \begin{tikzpicture}[node distance=1.5cm]

        \tikzstyle{startstop} = [rectangle, rounded corners, minimum width=2.5cm, minimum height=1cm, text centered, draw=black, fill=red!30]
        \tikzstyle{process} = [rectangle, minimum width=2.5cm, minimum height=1cm, text centered, draw=black, fill=orange!30]
        \tikzstyle{processMean} = [rectangle, minimum width=2.5cm, minimum height=1cm, text centered, draw=black, fill=yellow!30]
        \tikzstyle{processCube} = [rectangle, minimum width=2.5cm, minimum height=1cm, text centered, draw=black, fill=blue!30]
        \tikzstyle{processBadPixel} = [rectangle, minimum width=2.5cm, minimum height=1cm, text centered, draw=black, fill=green!30]
        \tikzstyle{processFlat} = [rectangle, minimum width=2.5cm, minimum height=1cm, text centered, draw=black, fill=purple!30]
        \tikzstyle{arrow} = [thick,->,>=stealth]

        \node (start) [startstop] {Start};

        \node (dark) [process, below of=start] {Master dark creation};

        \node (flat) [process, below of=dark] {Flat field correction};

        \node (flatreduction) [processFlat, right of=flat, xshift=-4.7cm] {\parbox{2.5cm}{\centering Flat field\\reduction}};
        \draw [arrow] (flatreduction.east) -- ++(0,0) |- (flat.west);

        \node (badpixel) [process, below of=flat] {Bad Pixels correction};

        \node (badpixelmap) [processBadPixel, right of=badpixel, xshift=1.75cm] {\parbox{2.5cm}{\centering Bad pixel\\mask creation}};
        \draw [arrow] (dark.east) -- ++(1.75,0) |- (badpixelmap.north);
        \draw [arrow] (badpixelmap.west) -- ++(0,0) |- (badpixel.east);

        \node (background) [processMean, below of=badpixel] {Background subtraction};

        \node (alignment) [process, below of=background] {PSF alignment};

        \node (darkhole) [processCube, below of=alignment] {Dark hole extraction};

        \node (frames) [process, below of=darkhole] {Frames selection};

        \node (sigma) [process, left of=darkhole, xshift=-1.75cm] {\parbox{2.5cm}{\centering PSF sigma\\estimation}};
        \draw [arrow] (alignment.west) -- ++(0,0) -| (sigma.north);
        \draw [arrow] (sigma.south) -- ++(0,0) |- (frames.west);

        \node (psfsubtraction) [processMean, below of=frames] {PSF subtraction};

        \node (end) [startstop, below of=psfsubtraction] {End};

        \draw [arrow] (start) -- (dark);
        \draw [arrow] (dark) -- (flat);
        \draw [arrow] (flat) -- (badpixel);
        \draw [arrow] (badpixel) -- (background);
        \draw [arrow] (background) -- (alignment);
        \draw [arrow] (alignment) -- (darkhole);
        \draw [arrow] (darkhole) -- (frames);
        \draw [arrow] (frames) -- (psfsubtraction);
        \draw [arrow] (psfsubtraction) -- (end);

    \end{tikzpicture}
    \caption{Flowchart of the data reduction procedure. Parallel processes are represented as side branches. Colors indicate the following: orange for operations applied to individual frames in the datacube, yellow for operations that collapse the entire cube into a single frame (mean or median), purple for operations involving two datacubes, and green for operations using standard calibration data provided by ESO.}
    \label{fig:flowchart}
\end{figure}

\section{vAPP PSF characterization} \label{PSFcarat}
\begin{figure}[h]
    \centering
    \includegraphics[width=0.45\textwidth]{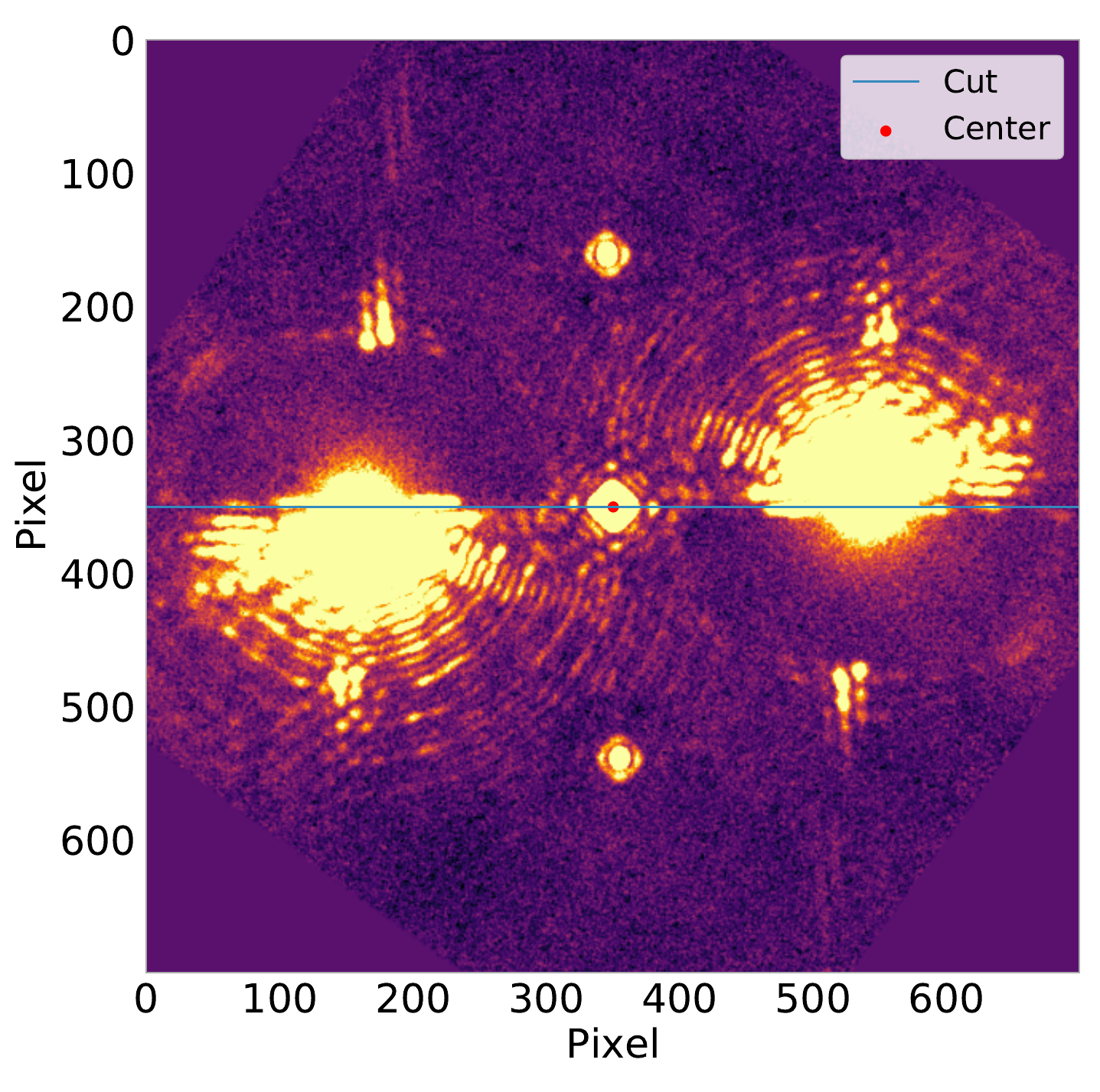}
    \caption{Rotated image of the reconstructed PSF by 36° with the rotation center highlighted and the position of the horizontal cut.}
    \label{TWHyapsfrotated}
\end{figure}
One crucial aspect of our work involved the characterization of the vAPP PSF. An initial step was to confirm that the vAPP was positioned at the specified $36^{\circ}$ angle on the telescope's pupil, as indicated by the instrument's manual. To achieve this, we rotated the PSF by $36^{\circ}$ and extracted a horizontal cut passing exactly through the center of the PSF (the leakage-term). To verify the angle, we ensured that the two first orders of the PSF on the right and left sides of the upper and lower PSF were symmetrical concerning their respective order 0. We used this extracted cut to fit three Gaussians (to model the order 0 of the PSF) at the upper, lower, and central positions. This process was performed for all targets, including the PSF reference reconstructed for RDI and the theoretical PSF.\\

\begin{figure}[h]
    \centering
    \includegraphics[width=0.5\textwidth]{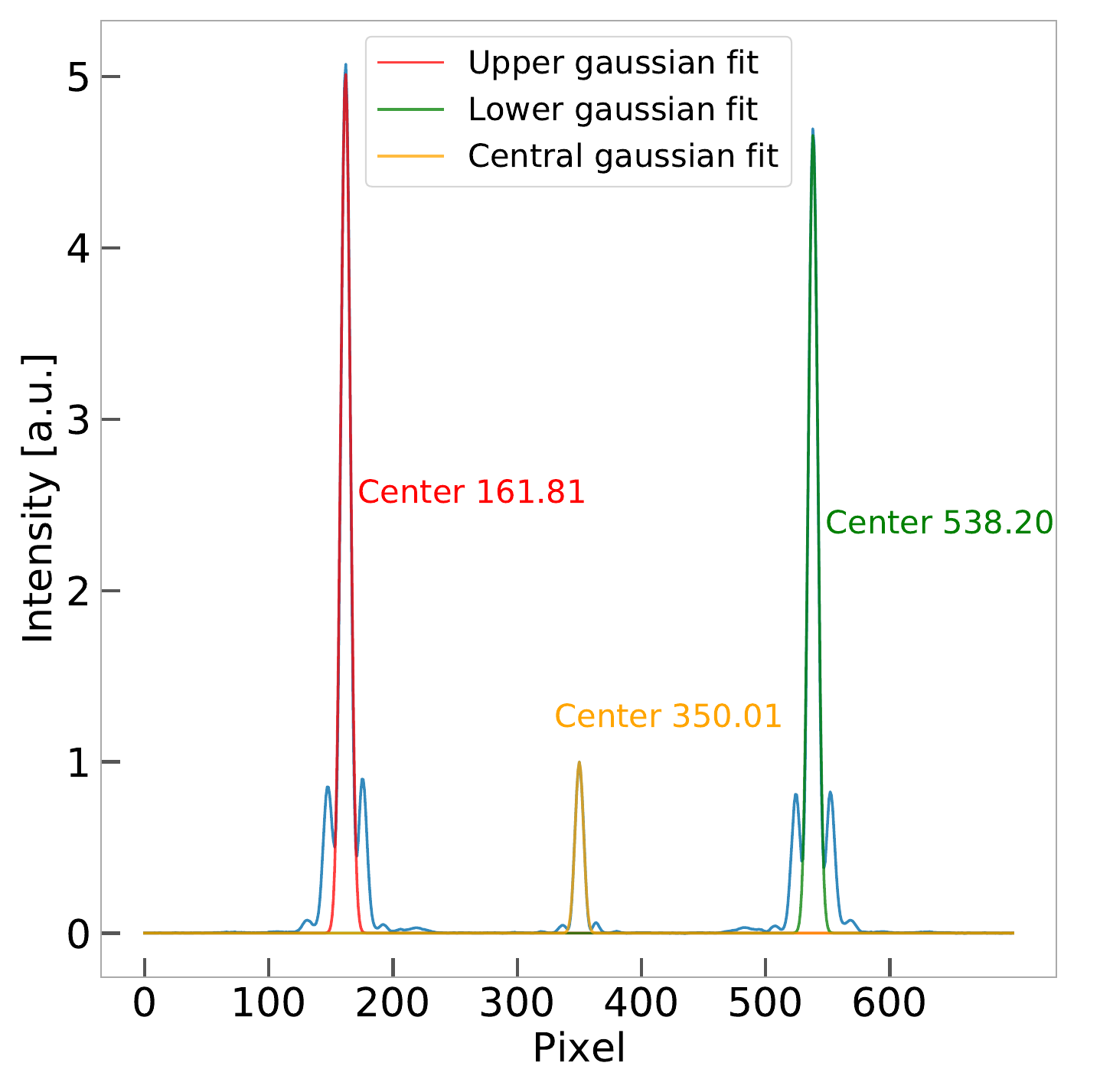}
    \caption{Horizontal cut of the vAPP PSF, with three Gaussian fits and the positions of the centers of the three Gaussians highlighted. The uncertainty is 0.01 for all measurements.}
    \label{gaussianfit3psf}
\end{figure}

\begin{figure}[h]
    \centering
    \includegraphics[width=0.5\textwidth]{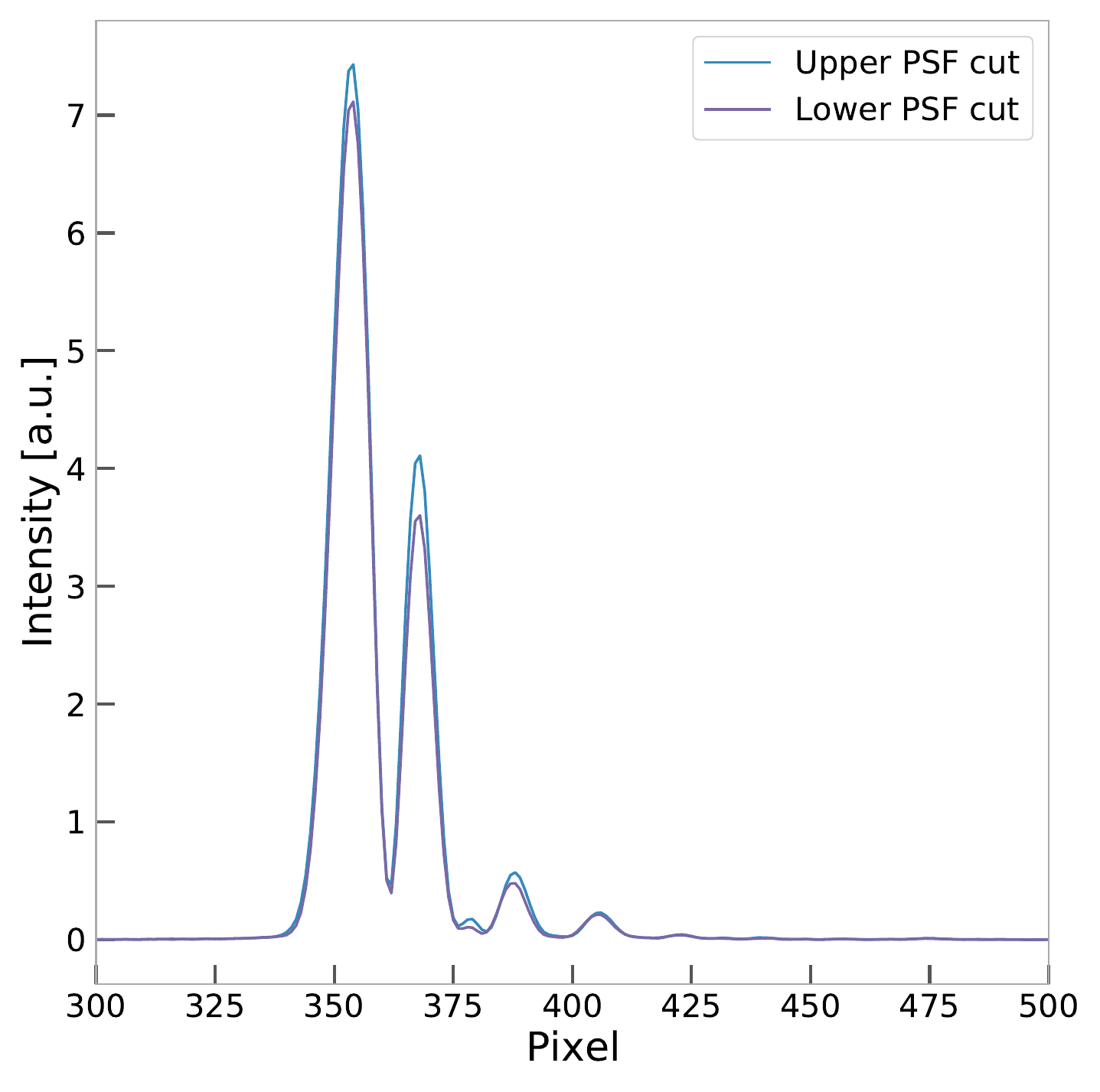}
    \caption{Comparison between the vertical cut of the PSF upper and lower. It is visible that the PSF lower, while having maxima at the same points, is significantly less luminous.}
    \label{verticalcut}
\end{figure}

\begin{figure}[h]
    \centering
    \includegraphics[width=0.5\textwidth]{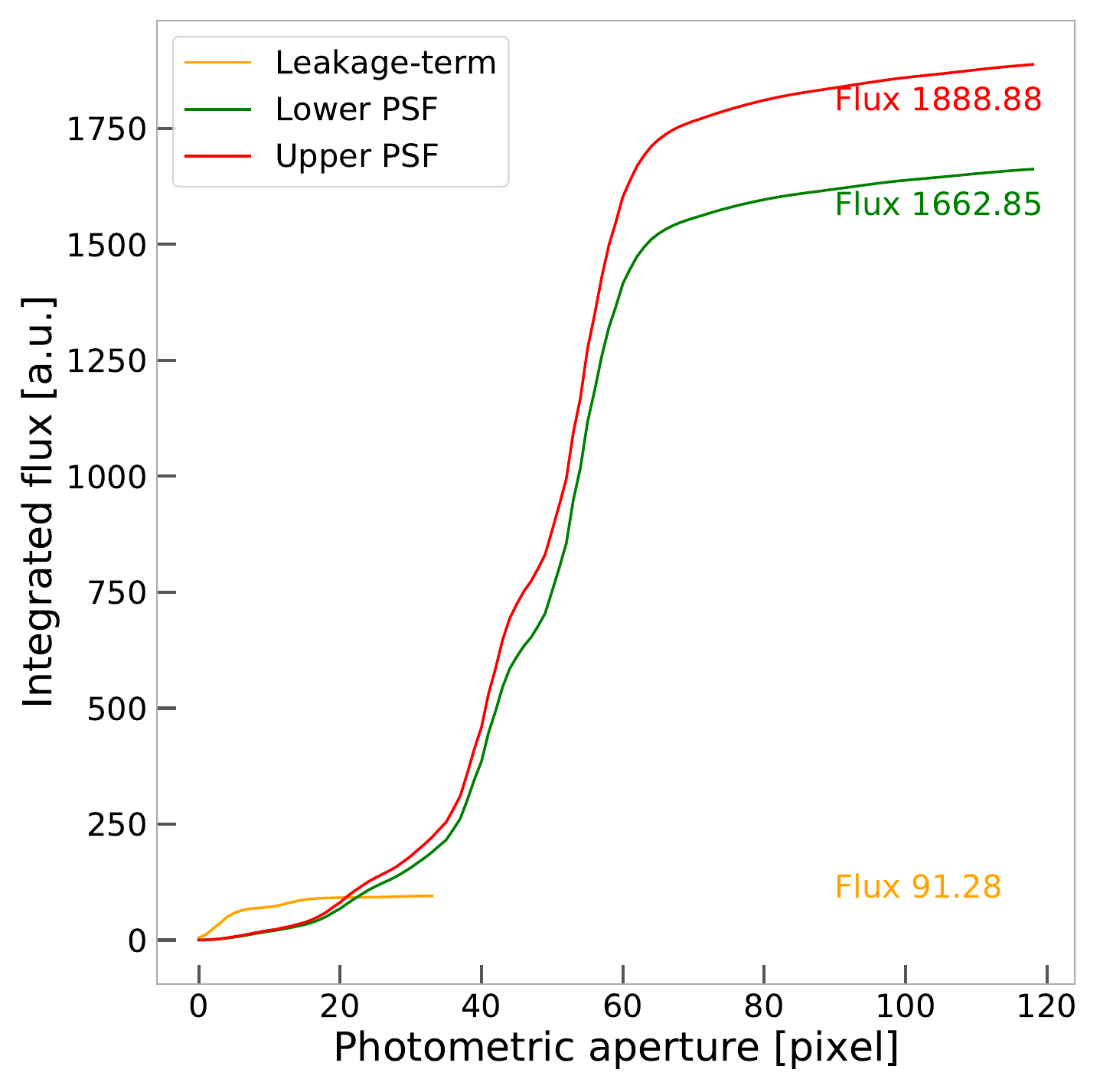}
    \caption{Flux variation in the PSF upper, lower, and leakage-term as a function of the photometric aperture.}
    \label{fotometriapsf}
\end{figure}

In Figure \ref{TWHyapsfrotated}, the rotated PSF of reconstructed is shown, with the highlighted line indicating where we extracted the cut. In Figure \ref{gaussianfit3psf}), the results of the three Gaussian fits are presented. Once the coordinates of the upper and lower PSF centers were determined, we made two additional cuts, vertically at the found coordinates. This allowed us to obtain a \enquote{radial} section of the PSF, as shown in Figure \ref{verticalcut}. Analyzing the shape and position of the various peaks of this PSF, which were compatible for both, revealed that the lower PSF was consistently less bright than the upper PSF. This constant flux asymmetry between the two vAPP PSFs suggests a possible instrumental effect, potentially linked to the presence of circular polarization induced by ERIS optics. To quantify this difference, we performed aperture photometry at variable radius for the two PSFs and the leakage-term (see Figure \ref{fotometriapsf}).

\begin{figure}[h!]
  \centering
    \includegraphics[width=0.22\textwidth]{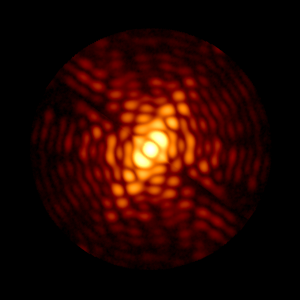}
    \includegraphics[width=0.22\textwidth]{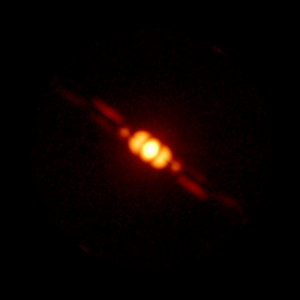}
    \includegraphics[width=0.22\textwidth]{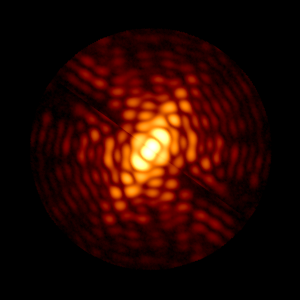}
    \includegraphics[width=0.22\textwidth]{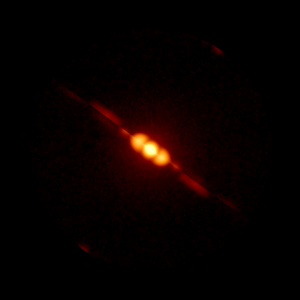}
  \caption{Upper line: Reconstruction of the union of the two bright sides of the PSF using the photocenter on the left and, on the right, the union of the two Dark Holes. Lower line: Reconstruction of the union of the two bright sides of the PSF using the rotation on the left and, on the right, the union of the two Dark Holes.}
  \label{PSF BS phot}
\end{figure}

Furthermore, having determined the centers of the PSFs, we could compare them with the photocenters of the upper and lower PSFs. This results show a clear difference between Photocenters and Rotation Centers. The rotation center is the point where the star should be, around which the elements in the dark Hole rotate. On the other hand, the Photocenter is the peak of brightness in the PSF. This represents a significant discontinuity compared to past works done with similar instruments to the ERIS vAPP \citep{Sutlieff.app}, where the Photocenter was always taken as a reference. This would lead to not rotating around the correct center in the ADI. In Figure \ref{PSF BS phot} first line, we show the PSF total brightness and the combination of the two dark Holes using the Photocenter as the center, while in Figures \ref{PSF BS phot} second line, the rotation center is used. It can be noticed that using the rotation center, the intrusions of the PSF into the dark hole are much less noticeable.\\

\begin{figure}[h!]
    \centering
    \includegraphics[width=0.5\textwidth]{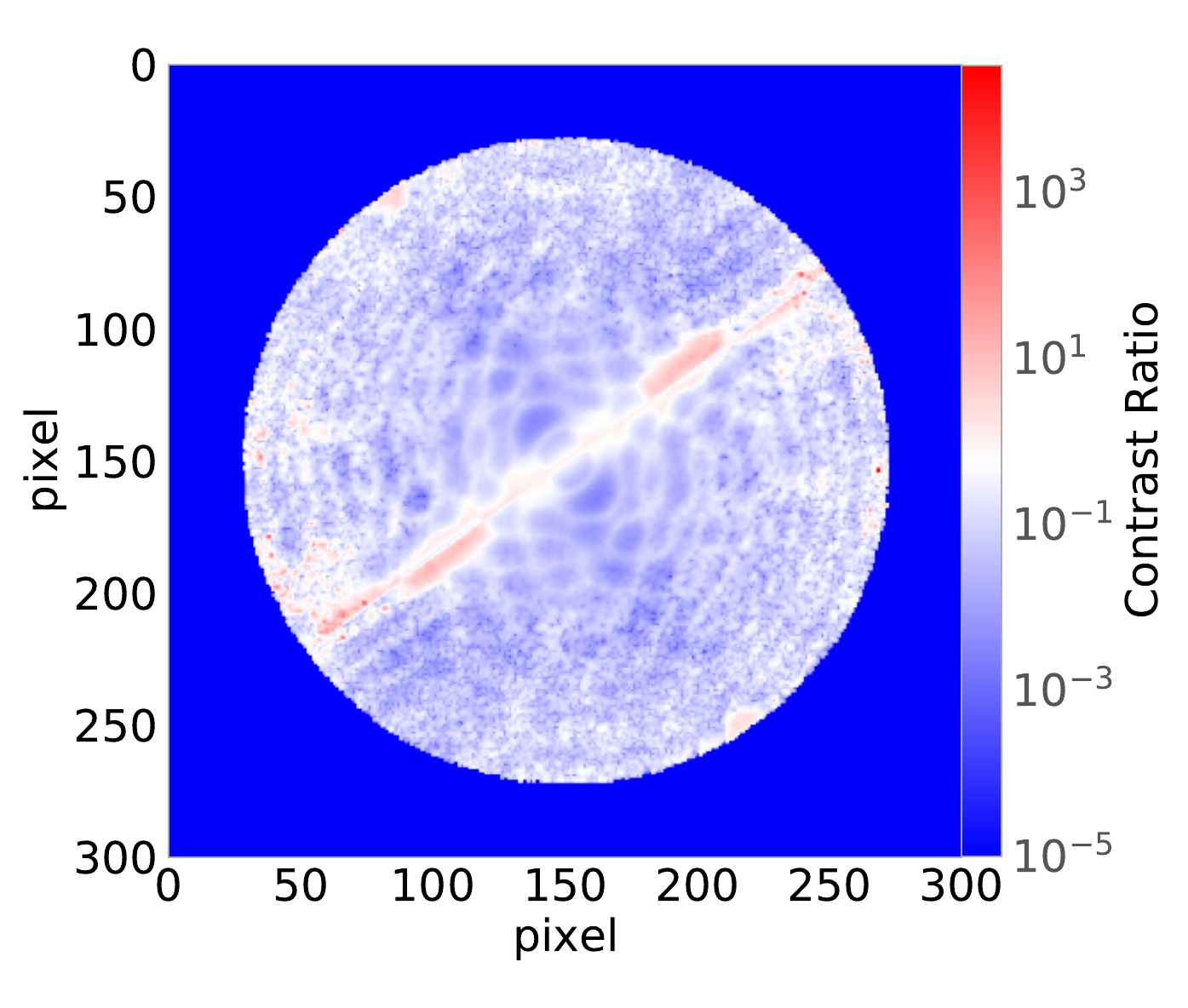}
    \caption{The map of the nominal contrast on the two Dark Holes shows that the worst contrast is obtained in the junction between the two.}
    \label{normcontrs}
\end{figure}

The ratio between the bright side and dark hole, in Figure \ref{PSF BS phot} second line, representing the Nominal contrast of the ERIS vAPP coronagraph without the use of High-Contrast Imaging (HCI) techniques, is shown in Figure \ref{normcontrs}. The mean value over the two dark Holes is around $7 \times 10^{-2}$, significantly lower than that achieved with cADI and ADI-PCA and estimated with the contrast curves.

This analysis is essential since all sources inside the dark holes are convolved with the same PSF. This may introduce artifacts around bright point-like and extended structures due to the asymmetry of the bright PSF of the vAPP. Careful interpretation of faint substructures is required.

\section{Role of photon noise in contrast curves}\label{PhotonNoise}
Analyzing the contrast curve computed in section \ref{contrast_curve_section} and shown in Figure \ref{contrasti} all reach a plateau at $\sim0\farcs4$, which corresponds to the observable contrast limit. Our analysis shows that this contrast is primarily driven by the photon noise of the background rather than PSF residuals.

To test this hypothesis, we selected a $500\times500$ pixel square region in the frame where the background was observed, corresponding to the expected position of the vAPP PSF. We estimated the mean counts per pixel in this area. The photon noise was computed as the square root of the mean counts. Since the background photon noise is present in both the science and background frames, we assumed them to be equal and combined them in quadrature.

\begin{figure}[h]
        \centering
        \includegraphics[width=0.45\textwidth]{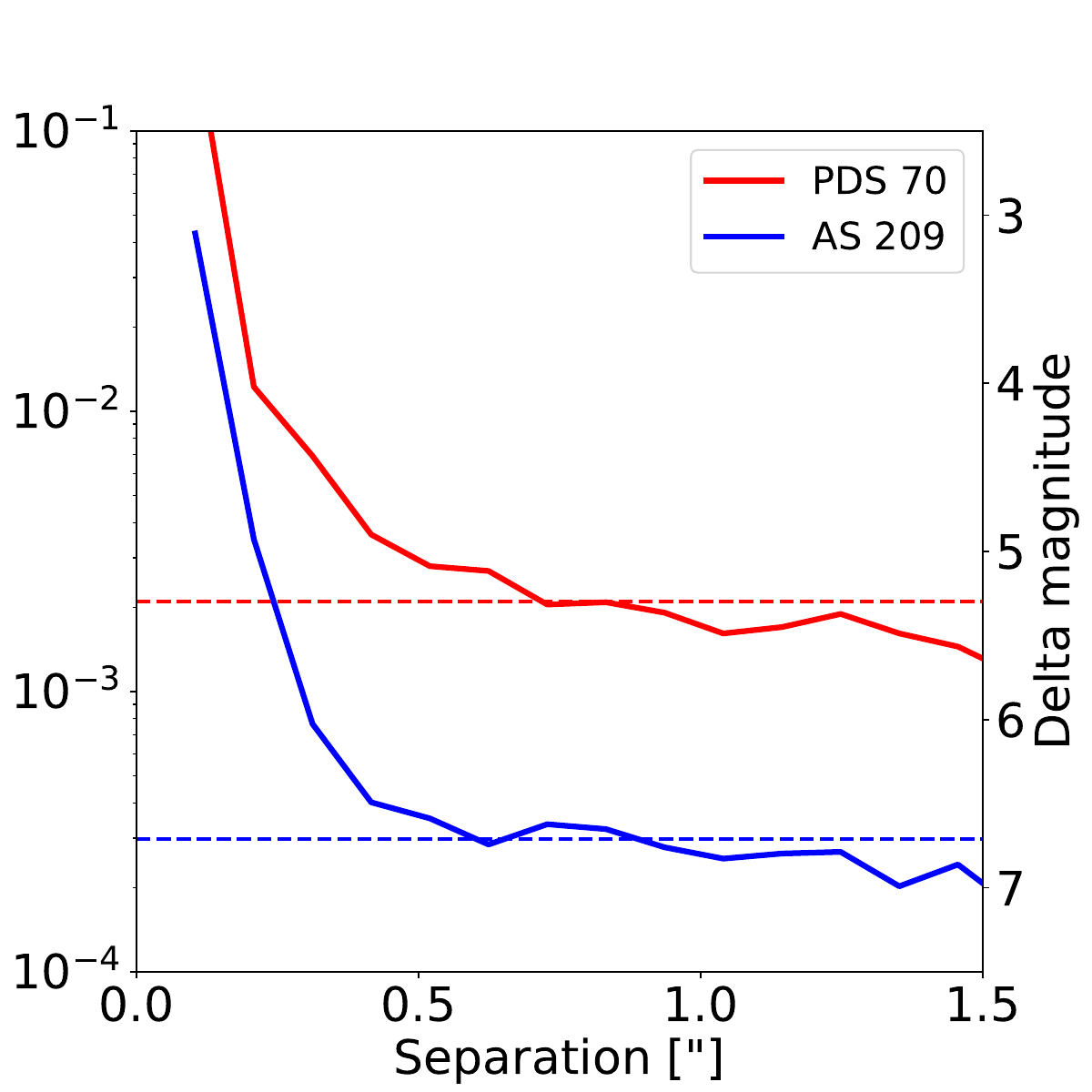}
    \caption{Example of a contrast curve comparison, PDS 70 (red) and AS 209 (blue). The solid line represents the contrast curve and the dotted line represents the contrast level of the photon noise.}
   \label{img:photnise}
\end{figure}

To estimate the contrast associated with this noise level, we used the photometry of the PSF leakage term. The result is consistent with the observed plateau in all contrast curves, see the example in Figure \ref{img:photnise}.

\section{Study of synthetic planets through transition zone} \label{fakeplanet-TZ}
The vAPP PSF is fixed with respect to the telescope pupil, remaining stationary during observations, while the field of view rotates. As a result, disk structures and planets can transition from a dark hole to the corresponding bright side of the PSF or vice versa in the opposite half of the PSF. When merging the two dark holes, this transition appears as a passage from one dark hole to the other. Consequently, a region of the field of $360^\circ$ view around the star, centered on the junction line of the two dark holes, undergoes this transition during the observation. We refer to this region as the transition zone, which extends: 

\begin{equation}  
DZ = \frac{FR}{2}
\end{equation}  

from both sides of the junction. Here, \(DZ\) represents the half angular extent of the transition region, and \(FR\) is the total field rotation of the observation. 

To assess the impact of this transition on observations, we analyzed a reference dataset: PDS 70 observations with a total field rotation of \(38^\circ\). We injected six synthetic planets, all with the same contrast, at \(1\farcs0\) and four at \(0\farcs25\). Two of the planets were positioned precisely on the dark hole junction to ensure they transitioned between regions during the observation sequence.

The synthetic planets were injected into the vAPP PSF dark holes as scaled copies of the corresponding bright PSFs. Field rotation was simulated by shifting the planets' position along a circular trajectory across frames. The resulting data cube, including the injected sources, was processed following the same reduction steps as the science data and analyzed using cADI and ADI-PCA.

\begin{figure}[h]
        \centering
        \includegraphics[width=0.45\textwidth]{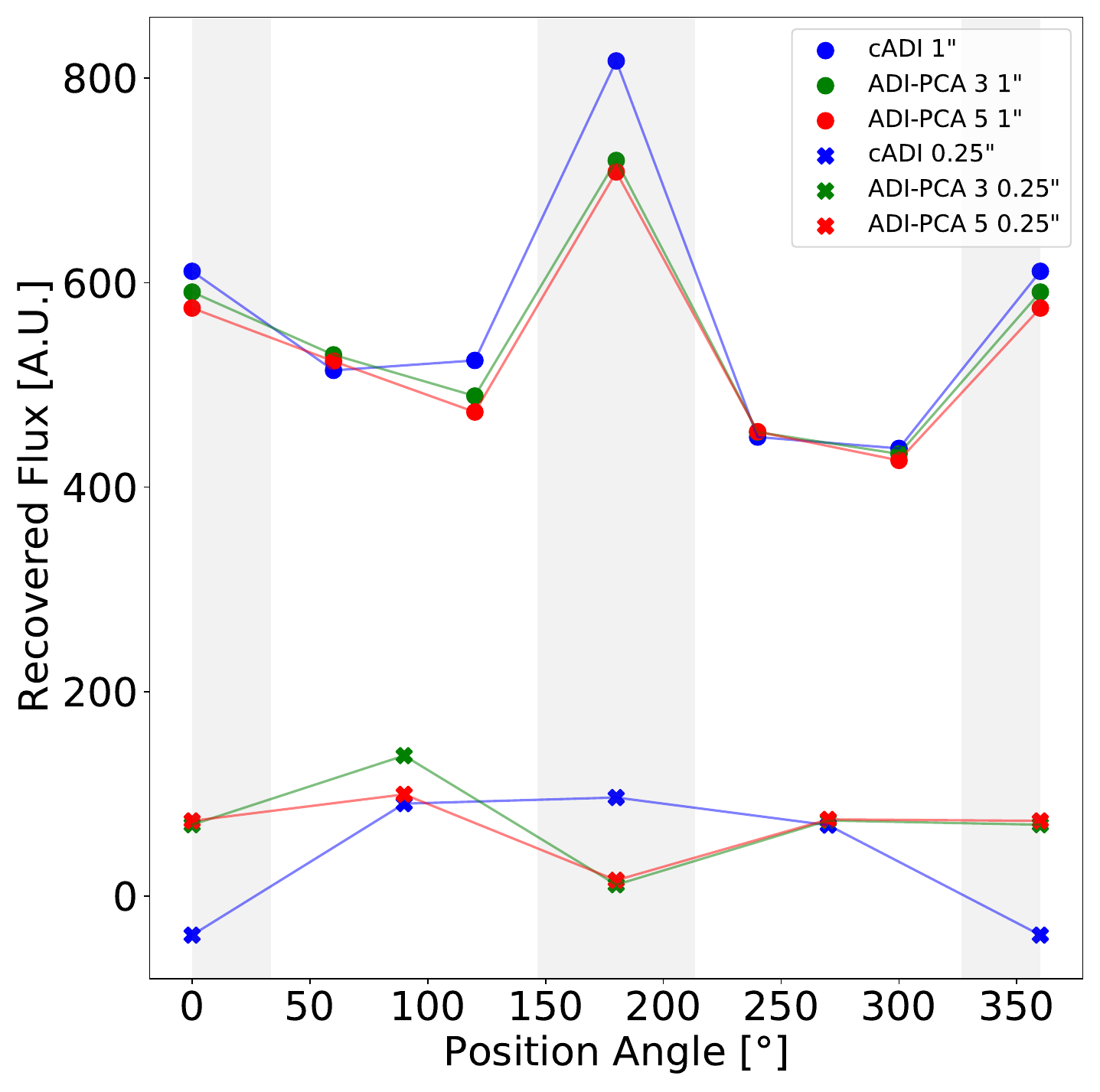}
    \caption{Recovered flux from aperture photometry for the six outer synthetic planets ($1\farcs0$) and the four inner planets ($0\farcs25$). The vertical gray bands indicate the transition regions for the PDS 70 observations. The colors represent different analysis techniques: blue for cADI, green for ADI-PCA with three subtracted PCs, and red for ADI-PCA with five subtracted PCs.}
   \label{img:posizione_pianeti}
\end{figure}

Planet fluxes were extracted via aperture photometry. As expected, due to the throughput dependence on angular separation from the central star, the recovered flux of inner planets was significantly lower than that of outer planets (see Figure~\ref{img:posizione_pianeti}). Notably, planets located in the transition regions (highlighted in gray in Figure~\ref{img:posizione_pianeti}) appeared systematically brighter at large angular separations and systematically fainter at small separations. The only exception was an inner planet in the cADI reduction.

\section{Background test} \label{Bactest}
\begin{figure*}[h]
        \centering
        \includegraphics[width=1.05\textwidth]{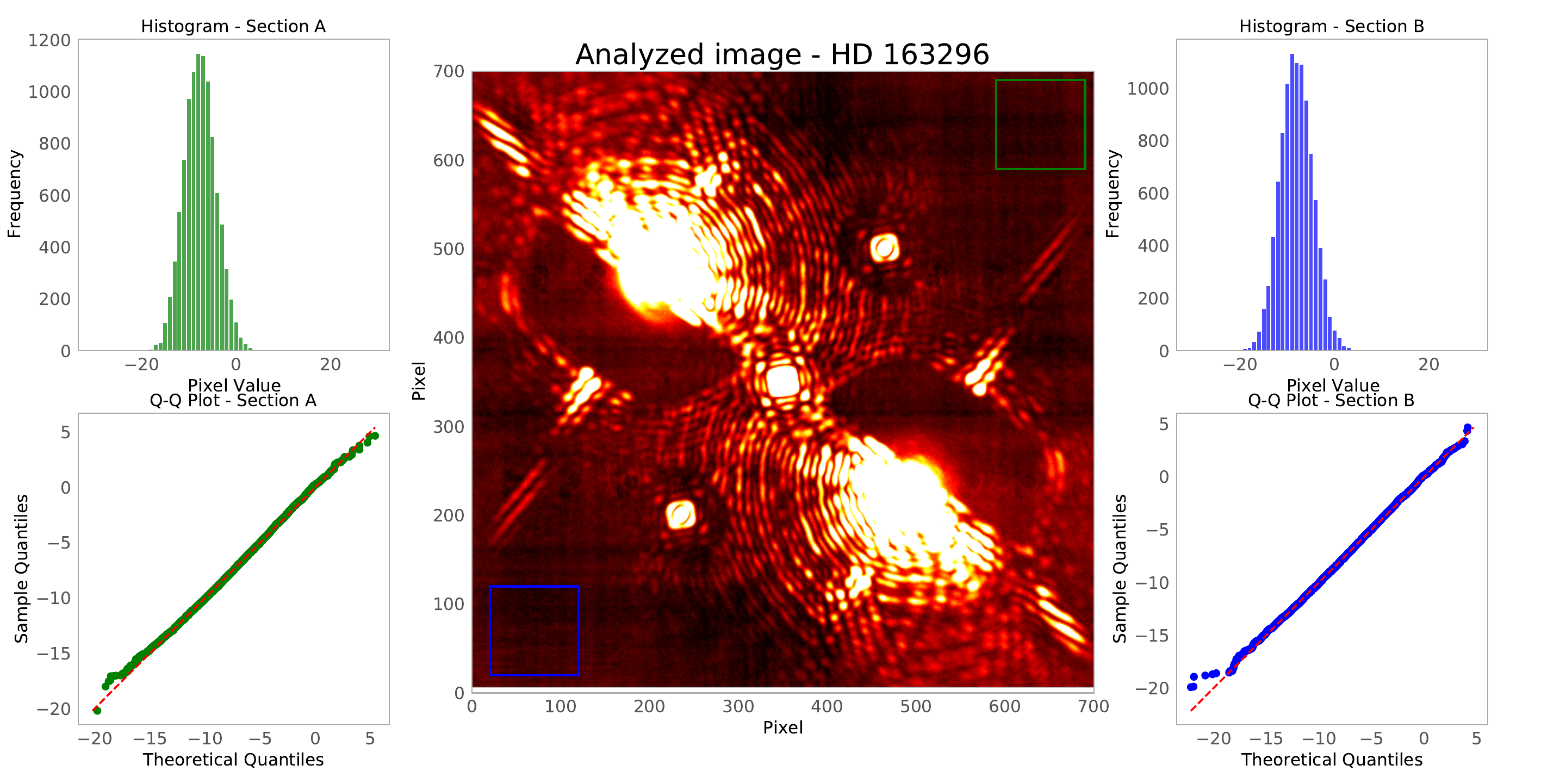}
    \caption{Histogram of the counts from two background portions of the image before dark hole extraction. In this example, a negative offset is present due to the occurrence of numerous negative bands, and the mean is non-zero.}
   \label{qq2}
\end{figure*}

\begin{figure*}[h]
        \centering
        \includegraphics[width=1.05\textwidth]{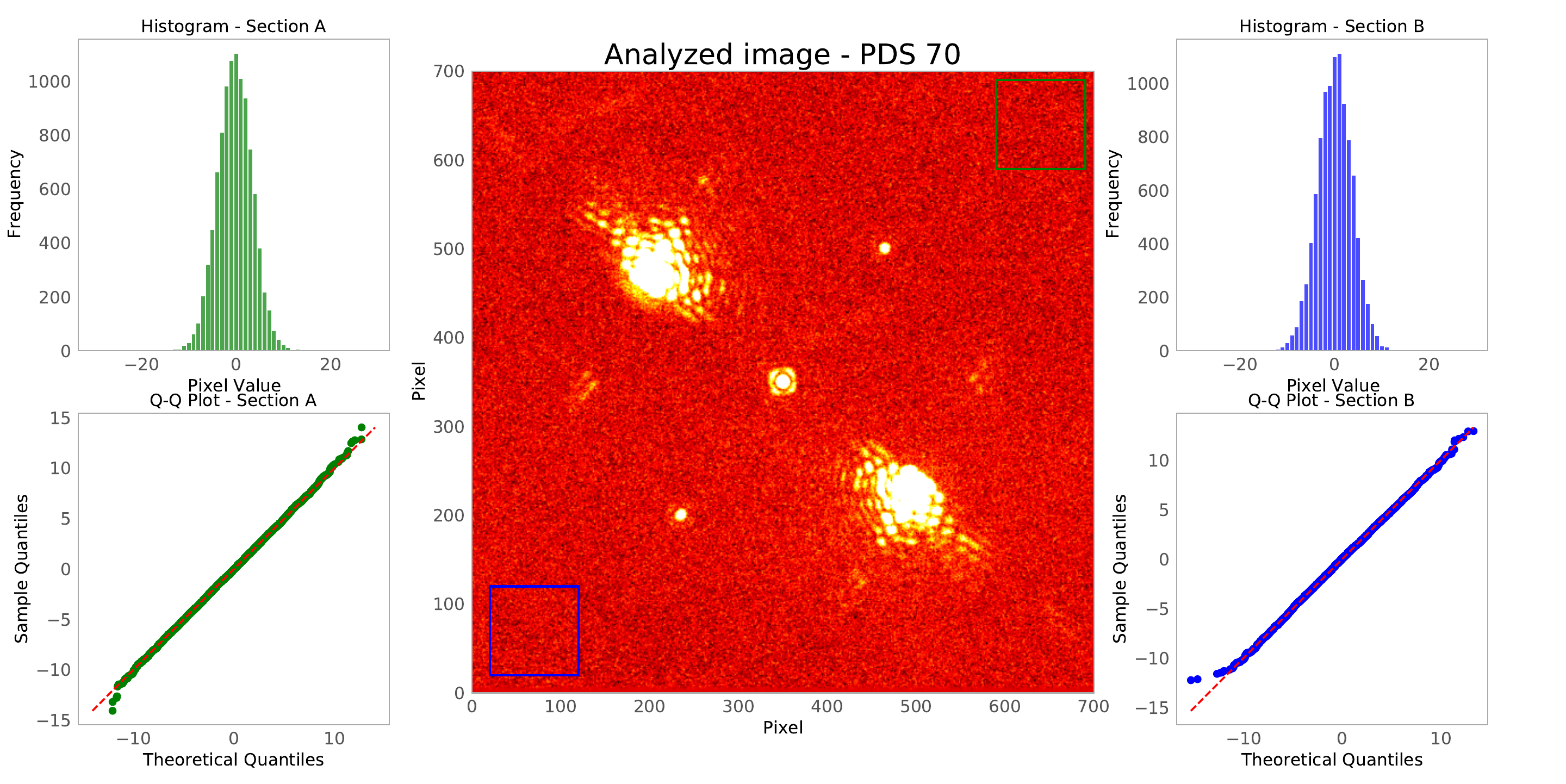}
    \caption{Histogram of the counts from two background portions of the image before dark hole extraction. In this example, the behavior is Gaussian, and the Q-Q plots are linear.}
   \label{qq1}
\end{figure*}

Once the reduction of the images was completed we performed tests to evaluate the behavior of the residual background. In particular, two tests were carried out before extracting the dark holes: the first was to analyze the distribution of counts in two areas of the image testing portions of PSF using of Q-Q plots (Figure \ref{qq2} and \ref{qq1}); the second test was the analysis of the value of the columns and rows of the image to evaluate the presence of structures such as bands remaining in the images. 

\begin{figure}[h]
        \centering
        \includegraphics[width=0.5\textwidth]{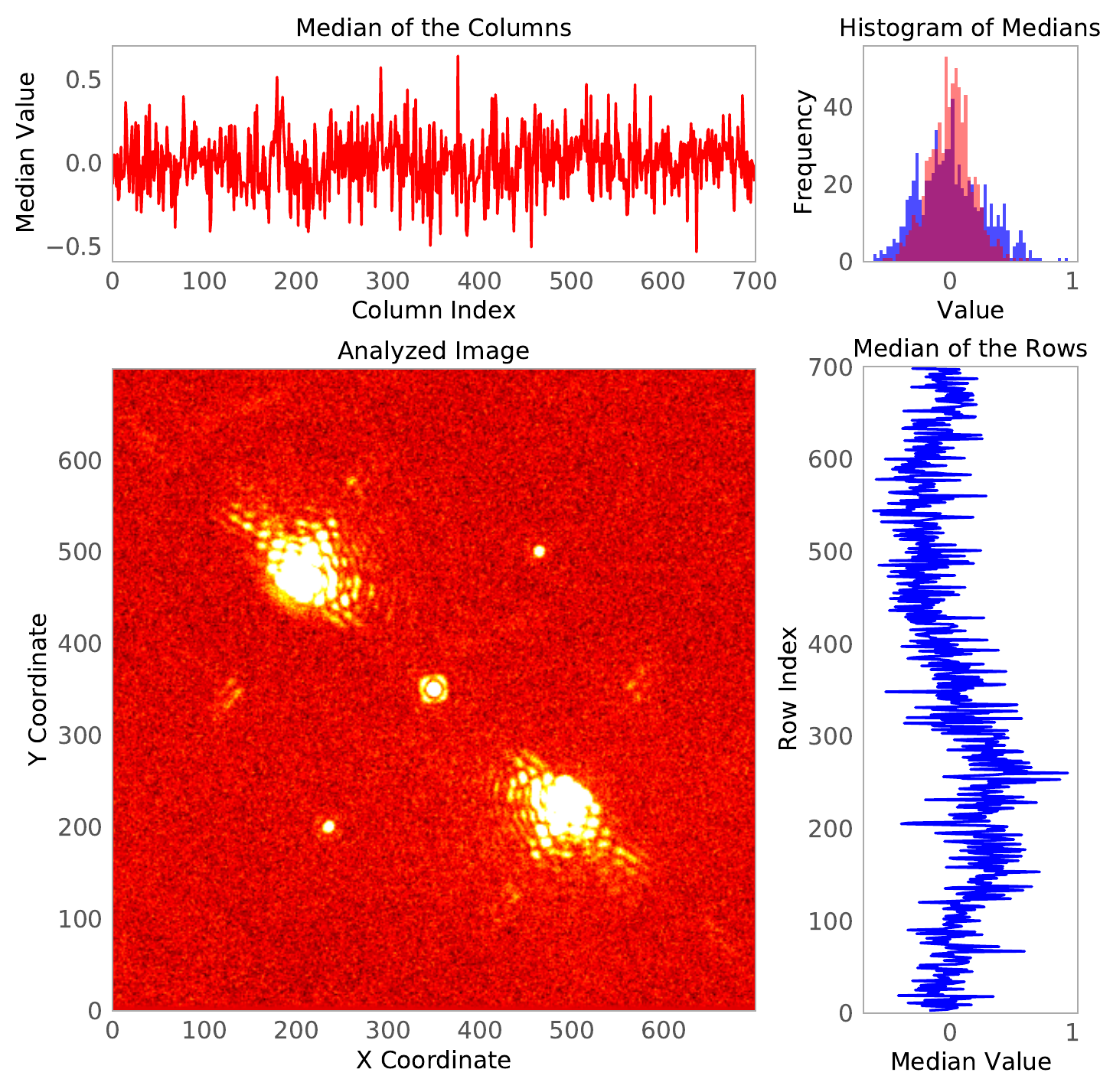}
    \caption{Example of the analysis of the vertical and horizontal stripes of the background. On the right you can find the trend of the counts of the lines and at the top of the columns. This example shows PDS 70 where the cleaning has left no obvious residues.}
   \label{bandeno}
\end{figure}

\begin{figure}[h]
        \centering
        \includegraphics[width=0.5\textwidth]{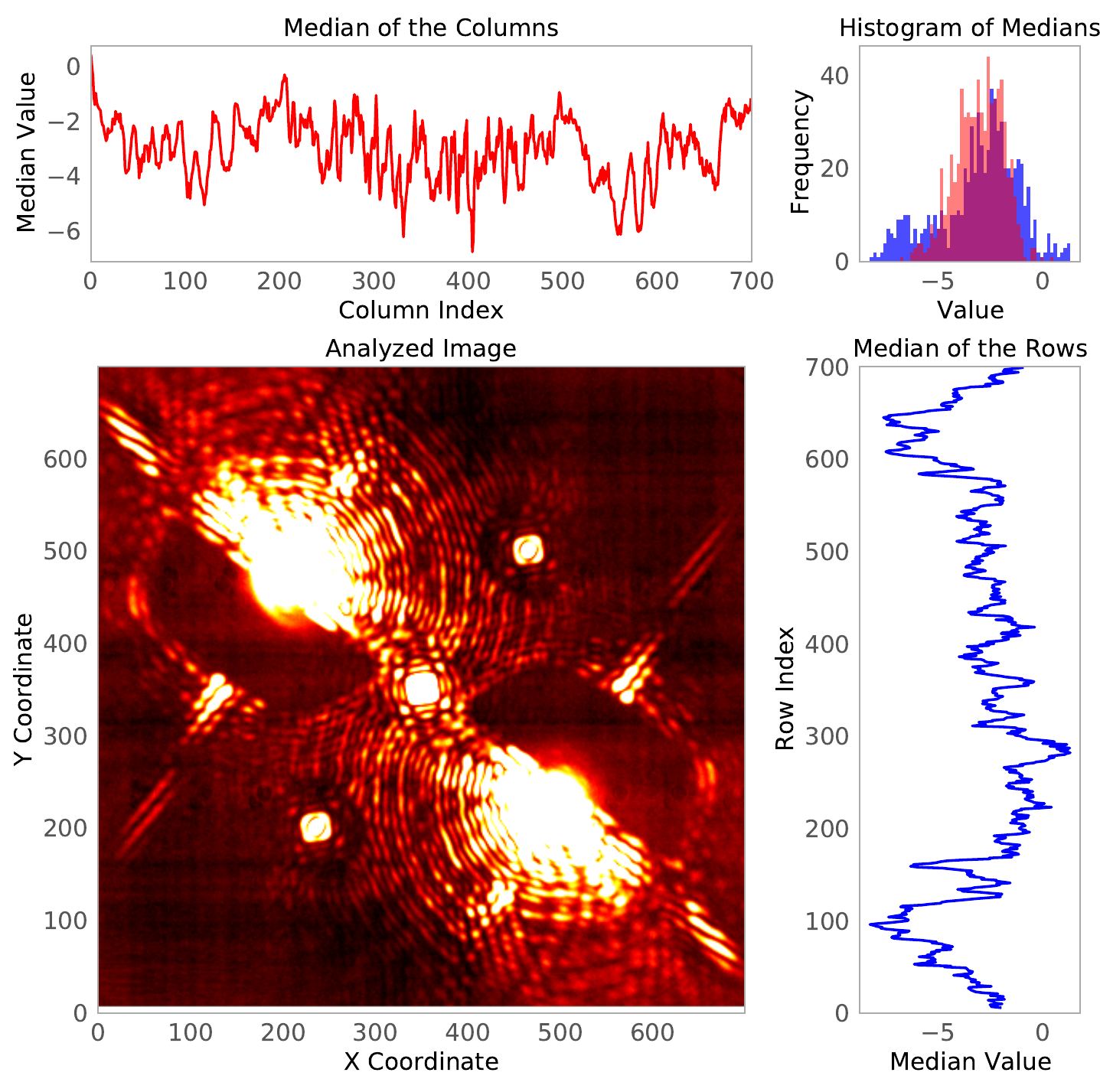}
    \caption{Example of the analysis of the vertical and horizontal stripes of the background. On the right you can find the trend of the line counts and at the top of the columns. This example shows HD 163296 where there are vertical and horizontal stripes left due to the saturation of the PSF.}
   \label{bandesi}
\end{figure}

There are both vertical and horizontal bands in the case of HD 163296 and HD 100546, for example, image \ref{bandeno} and \ref{bandesi}. The cause of these bands can be attributed to the extended and structured PSF of the APP: when the median of the rows and columns was taken it was hard to mask the entire PSF and leave enough \enquote{empty} background pixels and this caused an increase in the value of the median for that particular row or column. The subtraction of this value leads to further negative stripes.\\

The analysis of the Q-Q plot compares the quartiles of observed data with those of the theoretical distribution to visually assess the normality of HD 163296. The noise was locally Gaussian, though it was not centered around zero. However, by analyzing the plot of the median of rows and columns, in Figure \ref{qq2}, large oscillations across the entire field were present. These oscillations also explain the non-zero mean of the local noise distribution. This could significantly impact the analyses through RDI.\\

\newpage
\clearpage
\section{Temperature limit plot} \label{Temperaturelimitplot}

\begin{figure}[h!]
    \centering
    \includegraphics[width=0.8\textwidth]{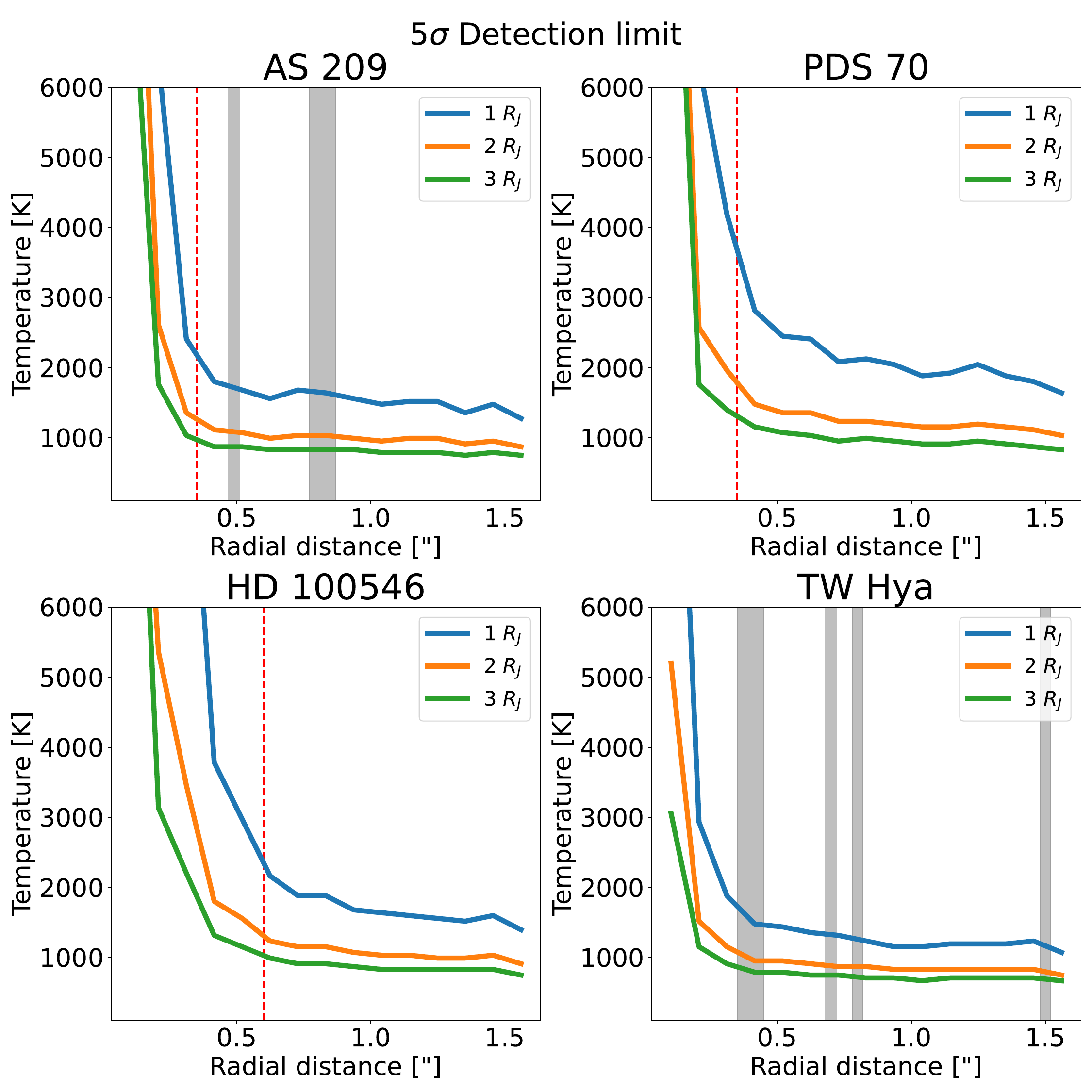}
    \caption{Minimum detectable temperature plots for the three selected planetary radii, shown for stars with rotation angles greater than 10 degrees. The vertical gray bands represent the position of gaps in the disk, while the vertical dashed red line indicates the angular distance within which disk residuals influenced the contrast curve.}
    \label{TEMP_LIM}
\end{figure}
\end{appendices}

\end{document}